\documentstyle[prd,aps,preprint,tighten,eqsecnum,epsf]{revtex}

\newcommand{\bbE}{\bar{\boldmath \mbox{$E$}{\,}}{\!}}
\newcommand{\bbJ}{\bar{\boldmath \mbox{$J$}{\,\,}}{\!\!}}
\newcommand{\bbN}{\bar{\boldmath \mbox{$N$}{\,}}{\!}}
\newcommand{\bbPi}{\bar{\boldmath \mbox{$\Pi$}{\,}}{\!}}
\newcommand{\boldeta}{{\boldmath \mbox{$\eta$}{\,}}{\!}}
\newcommand{\sigL}{\sigma_{\!{\scriptscriptstyle L}}}
\newcommand{\sigK}{\sigma_{\!{\scriptscriptstyle K}}}
\newcommand{\sigw}{\sigma_{\!{\scriptscriptstyle W}}}
\newcommand{\sigplus}{\sigma_{+}}
\newcommand{\sigminus}{\sigma_{-}}
\begin{document}
\preprint{TUW-96-27 and {\tt gr-qc/9612021}}

\title{Boundary Conditions and Quasilocal Energy in 
the\\Canonical Formulation of All $1+1$ Models of 
Gravity\footnote{Research supported by the ``Fonds zur 
F\"{o}rderung der wissenschaftlichen Forschung'' in Austria.}}

\author{W. Kummer and S. R. Lau\footnote{email addresses: 
{\tt wkummer@tph.tuwien.ac.at} and 
{\tt lau@tph16.tuwien.ac.at}}}

\address{Institut f\"{u}r Theoretische Physik,\\
Technische Universit\"{a}t Wien,\\
Wiedner Hauptstra\ss e 8-10,\\
A-1040 Wien, \"{O}sterreich}

\maketitle
\begin{abstract}
Within a first-order framework, we comprehensively examine the 
role played by boundary conditions in the canonical formulation
of a completely general two-dimensional gravity model. Our 
analysis particularly elucidates the perennial themes of mass and 
energy. The gravity models for which our arguments are valid 
include theories with dynamical torsion and so-called 
generalized dilaton theories ({\sc gdt}s). Our analysis of the 
canonical action principle (i) provides a rigorous 
correspondence between the most general first-order 
two-dimensional Einstein-Cartan model ({\sc ecm}) and {\sc gdt} and 
(ii) allows us to extract 
in a virtually simultaneous manner the ``true degrees of freedom'' 
for both {\sc ecm}s and {\sc gdt}s. 
For all such models, the existence of an 
absolutely conserved (in vacuo) quantity $C$ is a generic 
feature, with (minus) $C$ corresponding to the black-hole mass 
parameter in the important special cases of spherically 
symmetric four-dimensional general relativity and standard 
two-dimensional dilaton gravity. The mass $C$ also includes 
(minimally coupled) matter into a ``universal mass function.''  
We place particular emphasis on the (quite general) class of 
models within {\sc gdt} possessing a Minkowski-like groundstate 
solution (allowing comparison between $C$ and the 
Arnowitt-Deser-Misner mass for such models).
\end{abstract}
\vfill
Vienna, December 1996 
\newpage


\section{Introduction and Preliminaries}


That black-hole physics plays a basic role in understanding 
the relationships between quantum-mechanics, thermodynamics,
and gravity is, by now, undisputed. 
It is also widely believed that the ``correct'' 
notion of gravitational (plus matter) energy is one basic 
stepping stone towards a deeper understanding of such 
interconnections. Now, in fact, a plethora of interesting and 
useful definitions of gravitational ``energy'' and ``mass'' 
exist\cite{manypeople}. Although many of these proposals are
quite general, in practice the first test-solutions for which 
a given proposal is evaluated are the spherically symmetric 
Schwarzschild and Reissner-Nordstr\"{o}m black holes ({\sc sbh} and 
{\sc rnbh}). Therefore, ``spherically symmetric general relativity''
({\sc ssgr}) continues to provide an important arena for examining the 
notion of gravitational energy. {\sc ssgr} is effectively a 
two-dimensional 
theory\cite{Thomi_Isaak_Hajicek,Kastrup_Thiemann,Kuchar}, as the 
angular coordinates can often be ``thrown away,'' leaving the 
remaining time and radial dimensions $(t,r)$ as the only 
coordinates of importance. In this regard {\sc ssgr} is similar to 
a generic 1+1-dimensional covariant gravity ``model''
\cite{Schaller_Strobl,Kloesch_Strobl,Kummer_Widerin,KKL1}, which 
suggests that there are in effect many ``testing grounds'' where 
one can gain experience and insight (which might be brought to 
bear later on the above mentioned issues). Along this line of 
research, it would seem quite important not to limit the 
collection of 1+1 models {\em a priori}, as particular models may 
not mimic certain important features of {\sc ssgr}. In particular, the 
currently popular two-dimensional dilaton gravity 
({\footnotesize 2d}{\sc dg})\cite{Sengupta_Wadia,Witten,CGHS} possesses a dilaton 
black-hole ({\sc dbh}) solution which differs physically from the {\sc sbh} 
in a potentially severe way\cite{KKL2}.  

Comparing with such grand intentions, we should point out
that the purpose of this paper is more modest.
Within a first-order framework, we comprehensively 
examine the role played by boundary conditions 
in the canonical formulation of a completely general 
two-dimensional gravity model. Our analysis particularly 
elucidates the perennial themes of mass and energy.  The 
definition of energy that we use corresponds to the 
one\cite{Brown_York} derived for full general relativity by 
Brown and York via a Hamilton-Jacobi-type analysis of the 
gravitational action, while the mass definition we study 
corresponds to a plethora of equivalent mass definitions in {\sc ssgr}
\cite{manypeople,Kastrup_Thiemann,Kuchar,Misner_Sharp,manypeople2}.
The gravity models for which our arguments are valid include 
theories with dynamical torsion and so-called generalized dilaton 
theories ({\sc gdt}s)\cite{GDTs}. 
Our analysis of the canonical action (i) provides a rigorous 
correspondence between the most general 2d
Einstein-Cartan model ({\sc ecm}) and {\sc gdt}\footnote{Thereby 
complementing the covariant argument\cite{KKL1} 
which establishes this equivalence. 
However, our approach is more rigorous as we pay strict 
attention to the important role played by boundary 
conditions in the equivalence.} and (ii) allows us to extract 
in a virtually simultaneous manner the true degrees of freedom 
for both the (vacuum) {\sc ecm}s and {\sc gdt}s we study. 
For all such models, the existence of an absolutely 
conserved (in vacuo) quantity $C$ is a generic feature, 
with (minus) $C$ corresponding to the black-hole mass parameter 
in the important special cases of spherically symmetric 
four-dimensional general relativity {\sc ssgr} and 
{\footnotesize 2d}{\sc dg}\cite{Kummer_Widerin}. The mass $C$ also includes 
(minimally coupled) matter into a 
``universal mass function.''
   (While we do not make a detailed examination 
   of a particular coupled gravity-matter system 
   in this paper, many of our results are 
   unaffected by the inclusion of minimally 
   coupled matter. Throughout the course of our 
   discussion, we point out precisely when and why 
   we assume the vacuum case.)  
We place particular 
emphasis on the (quite general) class of models within {\sc gdt} 
possessing a Minkowski-like groundstate solution [allowing 
comparison between $C$ and the Arnowitt-Deser-Misner 
({\sc adm}) mass\cite{ADM} for such models].

In what remains of this first section we collect our basic 
notations and geometric conventions, as well as summarize 
some essential material on which the present work is based. 
In $\S$ II we discuss the kinematic relationship between
certain {\sc adm} decompositions of 
spacetime and two-dimensional moving frames ({\em zweibeine}). 
In $\S$ III we consider the general solution for the spacetime 
metric corresponding to the models we study, while in $\S$ IV 
we focus attention on those models which we consider to be 
``physical''. We carry out our investigation of gravitational 
energy and mass in the central $\S$ V, and the results of this section 
are applied in $\S$ VI. Finally, $\S$ VII offers some 
concluding remarks and outlook.

\subsection{Geometry of spacetime ${\cal M}$}
Consider a spacetime {\em region} ${\cal M}$ with boundary 
$\partial{\cal M}$. We put off a detailed description of 
${\cal M}$'s boundary structure until the next section. For 
the moment, it suffices to note that $\partial{\cal M}$ is 
composed of both timelike and spacelike elements. 
On ${\cal M}$ we have a zweibein $e^{a}{}_{\mu}$, with 
corresponding Lorentz-signature metric $g_{\mu\nu}$. 
Coordinate indices $(\mu,\nu,\cdots)$ run over $(t,r)$, 
zweibein indices $(a,b,\cdots)$ run over $(+,-)$, and the 
frame or ``internal'' metric is $\eta_{ab}$ (with $\eta_{+-}
= -1$ and $\eta_{++} = \eta_{--} = 0$). 
Let $\omega^{a}{}_{b\mu} =: -\epsilon^{a}{}_{b} \omega_{\mu}$ 
denote the coefficients (with respect to the zweibein) of the 
spacetime connection which, though compatible with the frame 
metric $\eta_{ab}$, is {\em not} necessarily torsion-free. 
Therefore, in general, $\omega^{a}{}_{b\mu}$ is built both 
from the zweibein $e^{a}{}_{\mu}$ and the spacetime torsion 
tensor $T^{\lambda}{}_{\mu\nu}$ (skew 
in its last two indices). We let $D_{\mu}$ represent the 
associated covariant derivative operator which ``sees'' zweibein 
indices, reserving $\nabla_{\mu}$ for the standard 
coordinate covariant derivative. Finally, the ${\cal M}$ 
permutation symbol is fixed by $\epsilon_{+-} = -1$, and the 
corresponding volume form is $\epsilon = \frac{1}{2}
\epsilon_{ab} e^{a} \wedge e^{b}$. 

\subsection{Action for an Einstein-Cartan model}
The action functional $L$ which describes our spacetime
geometry is the one appropriate for a general 
{\sc ecm}\cite{Schaller_Strobl,Kloesch_Strobl,Kummer_Widerin}, 
namely, a first-order action in terms of (what turn out
to be) Cartan variables,\footnote{We remark that the 
first-order {\sc ecm} action is an example of a Poisson-sigma 
model, although as we have written (\ref{PSMaction}), the 
underlying Poisson structure is not manifest as it is 
in Refs.~\cite{Schaller_Strobl,Kloesch_Strobl}. Our 
action is far from the most general Poisson-sigma 
model. Nevertheless, it suffices for our general 
investigation of 1+1 gravity models. For a detailed
treatment of and unifying approach for 1+1 gravity based
on the Poisson-sigma-model formulation, 
see Refs.~\cite{Kloesch_Strobl} by Kl\"{o}sch and Strobl.} 
\begin{equation}
   L = 
  \kappa
  \int_{\cal M} (- X_{a} D e^{a} 
  + X {\rm d}\omega 
  - V \epsilon) + L_{\partial{\cal M}} + L^{(m)} - 
  L|^{\scriptscriptstyle 0}
{\,} ,
\label{PSMaction}
\end{equation}
where for convenience\footnote{We shall unapologetically 
switch back and forth between the index and abstract notation 
in order to use the best notation when it is fitting. As a rule,
we reserve the abstract notation for differential forms
and linear operations. Thus, for example, we write $e^{+}$, 
$\omega[n]$, and $n[X]$, respectively, in place of 
$e^{+}{}_{\mu}{\rm d}x^{\mu}$, $\omega_{\mu} n^{\mu}$,
and $n^{\mu} \partial_{\mu} X$ (obviously in these examples
$n$ is a vector field and $X$ is a scalar function).
However, in order to avoid confusion, we use the index notation for 
vector fields, like $e_{+}{}^{\mu}$ and $n^{\mu}$; and, therefore,
(unless it appears in a linear operation as above) the plain letter
$n$ is always the $1$-form $n_{\mu}{\rm d}x^{\mu}$.}  
we have switched to the index-free
language of differential forms to write (\ref{PSMaction}). 
Let us provide a brief description of the various terms
found in (\ref{PSMaction}). First, $L^{(m)}$ represents a 
possible matter contribution to the action, and 
$L |^{\scriptscriptstyle 0}$ represents a 
possible {\em reference term} (an essentially arbitrary 
functional of boundary data which is fixed in the
variational principle)\cite{Gibbons_Hawking,Brown_York}. 
Such a reference
term does not affect the variational principle. For simplicity
we shall ignore the reference term until $\S$ VI, when its
presence is important for deriving the correct concept of 
``rest-frame energy.'' $L_{\partial {\cal M}}$ represents a 
boundary term which is included in the definition of $L$. In 
this section, we shall assume that
\begin{equation}
  L_{\partial {\cal M}} =
- \kappa \oint_{\partial {\cal M}}X\omega
{\,} ,
\label{naive_surface_term}
\end{equation} 
although later in $\S$ V we find the need to introduce an 
``improved'' (zweibein-gauge independent) version of 
this integral. Via an integration by parts we can 
replace the term $X{\rm d}\omega$ in the volume 
integral with $\omega \wedge {\rm d}X$, along the way 
generating a boundary term which exactly cancels the one 
that we have included. Next, $\kappa$ is a 
(possibly dimensionful) constant which allows us to easily 
compare our results with standard ones for {\sc ssgr} and 
{\footnotesize 2d}{\sc dg}. 
Finally, the potential term is
\begin{equation}
  V  =  
  \alpha X^{+} X^{-} 
+ \lambda^{2} V_{0}
{\,} ,
\label{PSMpotential}
\end{equation} 
where $\lambda$ is a constant with units of inverse length 
and $\alpha  = \alpha(X)$, $V_{0} = V_{0}(X)$. The 
``target-space coordinates''\cite{Schaller_Strobl,Kloesch_Strobl} 
$(X^{a},X)$ have transparent geometric meanings. 
The special case of {\sc ssgr} suggests that we may interpret $X$ 
as essentially the squared areal radius. 
Moreover, for {\sc ssgr} 
the $X^{a}$ turn out to be (apart from 
$X$-dependent factors) the expansions associated with the 
null zweibein. In addition to this special-case interpretation, 
the fields $X^{a}$ are also closely related to the spacetime torsion
when torsion is present ($\partial V/\partial X^{a} \neq 0$).
Indeed, among the Euler-Lagrange equations of 
motion ({\sc eom}s) associated with (\ref{PSMaction}) is the following:
\begin{equation}
  {\rm d}e^{a} - \epsilon^{a}{}_{b}\omega \wedge e^{b} =
  \alpha X^{a} \epsilon
{\,} .
\end{equation}
Therefore, one finds that $\alpha X^{a} = {}^{*} T^{a} :=
\frac{1}{2}\epsilon^{\mu\nu} e^{a}{}_{\lambda} 
T^{\lambda}{}_{\mu\nu}$ is the Hodge dual of the torsion
$2$-form.  

\subsection{Conformal transformation between physical and 
unphysical spacetimes}

In the absence of matter the {\sc eom}s associated with the {\sc ecm} 
action (\ref{PSMaction}) can be solved exactly for integrable $\alpha(X)$ 
and $v(X)$\cite{Schaller_Strobl,Kloesch_Strobl,Kummer_Widerin}. 
In this paper, (\ref{PSMaction}) (possibly also with matter) and 
its associated {\sc eom}s dictate the 
spacetime geometries we study, in particular determining
(what we call) the 
{\em unphysical} metric $g_{\mu\nu}$. However, we have reason to be 
interested in certain {\em physical} metrics $\tilde{g}_{\mu\nu}$ 
which arise from the unphysical metrics $g_{\mu\nu}$ via conformal 
transformation.
Our interest in the conformally related geometries stems from the 
local equivalence between a wide class of {\sc ecm} models (even many 
with non-vanishing torsion\cite{KKL1}) and so-called generalized 
dilaton theories ({\sc gdt}s)\cite{GDTs}. 
In the establishment of this equivalence,
such {\sc gdt}s emerge via the type conformal transformation we study here.
Now, in fact, both {\sc ssgr} and {\footnotesize 2d}{\sc dg} are special cases of {\sc gdt}.
Therefore, the conformal transformations we consider now are
essential, if we intend to apply our methods to (at least arguably) 
the most interesting 2d models. For example, we shall write down 
a {\sc ecm} action corresponding to vacuum {\sc ssgr}. However, in
this case it is the rescaled metric $\tilde{g}_{\mu\nu}$ rather than
the metric $g_{\mu\nu}$ (determined directly by this particular {\sc ecm} 
action's {\sc eom}s) which matches the familiar time-radial piece of the 
4d {\sc sbh} line-element\cite{Schaller_Strobl,Kloesch_Strobl,Kummer_Widerin}.

Our conformal transformations are given by the following set
of field re-definitions in the action (using abstract notation): 
\begin{eqnarray}
  e^{a} & = & \exp(\varphi) \tilde{e}^{a}
  \label{conformal} \eqnum{\ref{conformal}a}\\
  X^{a} & = & \exp(-\varphi) \tilde{X}^{a}
  \eqnum{\ref{conformal}b}\\
  \omega & = & 
  \tilde{\omega} - ({\rm d}\varphi/{\rm d}X){\,}
  {}^{\star}{\rm d}X
  \eqnum{\ref{conformal}c}
{\,} ,
\addtocounter{equation}{1} 
\end{eqnarray}
where $\varphi = \varphi(X)$ is at this point an arbitrary 
function of $X$ and the ``$\star$'' represents the Hodge-duality 
associated with physical metric $\tilde{g}_{\mu\nu}$. 
The transformation (\ref{conformal}) has been tailored to ensure 
that
\begin{equation}
T^{\lambda}{}_{\sigma\mu}  =  
  \tilde{T}^{\lambda}{}_{\sigma\mu}
\label{torsionunderconf}
\end{equation}
is a consistent behavior for the torsion tensor under the 
transformation. Under the conformal transformation 
(\ref{conformal}), the
ECT action (\ref{PSMaction}) becomes
\begin{equation}
   \tilde{L} = 
  \kappa \int_{\cal M} 
  \left[- \tilde{X}_{a} \tilde{D} \tilde{e}^{a}
  + X {\rm d}\tilde{\omega} + ({\rm d}\varphi/{\rm d}X) 
  {\rm d}X \wedge{\,} {}^{\star}{\rm d}X
  - \tilde{V} \tilde{\epsilon}\right]
- \kappa \oint_{\partial{\cal M}} X \tilde{\omega} + L^{(m)}
{\,} ,
\label{hatPSMaction}
\end{equation}
with $\tilde{V}$ determined by the previous transformation rule
(\ref{conformal}b) as well as $V_{0} = \exp(-2\varphi)\tilde{V}_{0}$. 
The boundary term that we have 
included in (\ref{PSMaction}) ensures that the variational 
principle associated with the rescaled action 
(\ref{hatPSMaction}) features fixation of the same 
boundary data as fixed in the original variational principle.
In the variational principle associated with (\ref{PSMaction}) 
$e^{a}{}_{\mu}$ and $X$ (and matter) are fixed, 
while in the variational principle associated with 
(\ref{hatPSMaction}) $\tilde{e}^{a}{}_{\mu}$ 
(which is built from $e^{a}{}_{\mu}$ and $X$) and $X$ (and matter)
are fixed. 
Preservation of the boundary conditions under the conformal
rescaling is the main motivation for including this boundary term,
although this requirement alone does not uniquely determine
the choice of boundary term.\footnote{Chan {\em et al} have 
carefully considered conformal transformations and the role 
played by quite similar boundary terms in general 
$n$-dimensional dilaton 
theories.\cite{Chan_Creighton_Mann} In $\S$ V we comment further
on the relationship between the terms we consider and those found
in that reference.} 
Indeed, in $\S$ V we shall add further boundary terms to the 
action which are insensitive to the conformal transformation. 

The form (\ref{hatPSMaction}) of the conformally rescaled action
affords a rather direct comparison with the second-order action 
associated with a {\sc gdt}. For simplicity here only, 
we restrict ourselves to
vanishing torsion. Indeed, with $\partial V/ \partial X^{a} = 0$
(no torsion), elimination of the fields $\tilde{X}^{a}$ via their
algebraic {\sc eom}s casts (\ref{hatPSMaction}) into
the following form:
\begin{eqnarray}
 \tilde{L} & = &
  {\textstyle \frac{1}{2}}
  \kappa \int_{\cal M} 
  {\rm d}^{2}x \sqrt{-\tilde{g}}
  [X \tilde{\cal R} - 2({\rm d}\varphi/{\rm d}X) 
  \tilde{g}^{\mu\nu}
   \partial_{\mu} X \partial_{\nu} X
- 2\lambda^{2} \exp(2\varphi) V_{0}]
\nonumber \\
& &
+ \kappa \oint_{\partial{\cal M}}{\rm d}x 
  \sqrt{|\tilde{g}^{1}|} 
  X \tilde{\epsilon}^{\mu\nu}
  \tilde{\omega}_{\mu}\tilde{\sf n}_{\nu} + L^{(m)}
{\,} ,
\label{hatdilatonaction}
\end{eqnarray}
where we have switched to index notation, $\tilde{\cal R}$ is the 
Ricci scalar of $\tilde{g}_{\mu\nu}$, and $\tilde{g}^{1}$ is the 
determinant of the induced metric on $\partial{\cal M}$ (the 
absolute value sign is needed to handle timelike elements of 
$\partial{\cal M}$). Further, $\tilde{\sf n}_{\mu}$ is the $1$-form 
dual (in the metric $\tilde{g}_{\mu\nu}$) to the outward-pointing 
unit-normal vector field of the boundary $\partial{\cal M}$ as 
embedded in ${\cal M}$ ($\tilde{\sf n}^{\mu}$ differs from the 
unit normal by a sign on a spacelike element of $\partial{\cal M}$). 
We note that, in general, the boundary 
term in (\ref{hatdilatonaction}) can not be expressed purely 
in terms of the metric $\tilde{g}_{\mu\nu}$ (rather, it is 
at heart a zweibein expression). Therefore, as {\sc gdt}s are, strictly
speaking, metric theories, the action above is not quite appropriate
to describe a {\sc gdt}. We shall improve upon this situation in $\S$ V,
where we begin with the addition of a boundary term to the {\sc ecm}
action which is zweibein-gauge invariant (in a sense to be made
precise below).

Let us quickly consider the special case of vacuum {\sc ssgr}, as it
will determine some of our conventions. Set $L^{(m)} = 0$ and 
make the following substitutions in (\ref{hatdilatonaction}): 
\begin{eqnarray} 
  \kappa^{\scriptscriptstyle SSGR} & = & 
  \lambda^{-2}
\label{SSGRchoices} \eqnum{\ref{SSGRchoices}a}\\
  V_{0}^{\scriptscriptstyle SSGR} & = &  
- \sqrt{{\textstyle \frac{1}{8}}} X^{-1/2}
\eqnum{\ref{SSGRchoices}b} \\
2\varphi^{\scriptscriptstyle SSGR} & = &
  \log \sqrt{2X}{\,} .
\eqnum{\ref{SSGRchoices}c}
\addtocounter{equation}{1}
\end{eqnarray}
Finally, set $X = {\textstyle \frac{1}{2}}\lambda^{2} R^{2}$, 
where $R(r,t)$ is the round $2$-sphere areal radius (or 
luminosity parameter). Notice that $X$ is restricted to positive 
values. The minus sign in (\ref{SSGRchoices}b) cancels one in
(\ref{hatdilatonaction}). We have introduced this spurious
minus sign only to ensure that our conventions are more
in harmony with those of Refs.~\cite{Kloesch_Strobl} (although 
those references do adopt a different metric-signature convention).
The result of these substitutions is an action which (as was first 
noted by Thomi, Isaak, and 
H\'{a}j\'{\i}\v{c}ek\cite{Thomi_Isaak_Hajicek}) has as its 
associated {\sc eom}s the 4d Einstein equations subject 
to the ansatz of spherical symmetry. In addition, the resulting 
action can (apart from certain technical points
concerning the zweibein dependence of the boundary term) be 
obtained directly form the 4d Einstein-Hilbert action via a 
reduction by spherical symmetry (see, for example, \cite{Lau}). 
The choices (\ref{SSGRchoices}) ensures that the {\sc adm} mass as 
measured at spatial infinity corresponds to the {\sc sbh} mass 
parameter. Moreover, the choice (\ref{SSGRchoices}a) gives the
action the units of action in 4d.

{\sc ssgr} is an example of the ``physical'' situation that we are 
interested in. Namely, we shall require that a flat line-element
(``Minkowski groundstate'') is obtained when the mass parameter
of the solution is set to zero. We consider the most general 
class of such models in $\S$ IV. It should be emphasized that the 
conformal transformations used in this paper leave minimal 
interactions with a scalar field invariant. A minimally coupled 
fermion transforms covariantly.

\subsection{The first-order approach}

We stress that we do {\em not} examine in detail the 
{\sc eom}s associated with the action (\ref{hatdilatonaction}), or, 
for that matter, those associated with the action (\ref{hatPSMaction}). 
Indeed, since these {\sc eom}s are considerably more complicated than 
those associated with (\ref{PSMaction}), it is prudent to work 
directly at the level of the {\sc ecm} action (\ref{PSMaction}) with 
its relatively simple first-order {\sc eom}s; switching via conformal 
transformation to the hatted variables of real interest only ``at 
the end of the day.'' The merit of this approach is strikingly
evident in the transparent derivation of the conservation law (always 
present for the 2d theories we study here)\cite{Kummer_Widerin}. 
Indeed, a simple 
linear combination of the {\sc eom}s obtained by varying the {\sc ecm} 
action (\ref{PSMaction}) with respect to the zweibein and 
spin connection yields the promised conservation law, namely,
\begin{equation}
{\rm d}[X^{+} X^{-}\exp(\alpha X) + W] + U^{(m)} = 0
{\, } ,
\label{conservation}
\end{equation} 
where for simplicity here we assume $\alpha$ is constant and
\begin{equation}
  W = 
  \lambda^{2} \int^{X}_{X_{0}}
  V_{0}(y) \exp (\alpha y) {\rm d}y
{\,} .
\label{W_definition}
\end{equation}
In the above conservation law $U^{(m)}$ arises from the 
presence of matter. Now, to make (\ref{conservation}) look like
a conservation law, we appeal to the Poincar\'{e}-Lemma which tells
us that that $U^{(m)} = {\rm d}C^{(m)}$ (at least locally), with 
$C^{(m)}$ a field-nonlocal expression built from the (geometric 
variables and) matter fields. That $U^{(m)}$ is locally exact must 
also follow via a clever combination of the remaining (notably 
matter) {\sc eom}s. 
Therefore, we have found
\begin{equation}
  {\rm d}[C + C^{(m)}] = {\rm d}[X^{+}X^{-}
  \exp(\alpha X) + W + C^{(m)}] = 0
\label{conservation2}
\end{equation}
for the conservation law valid for {\em all} {\sc ecm}s 
in 2d.\cite{Schaller_Strobl,Kloesch_Strobl,Kummer_Widerin} 
It is, of course, trivial to 
rewrite $C$ in terms of the physical variables. For the case 
of vacuum  {\sc ssgr} we shall see that $M_{\scriptscriptstyle S} = 
- \lambda^{-3} C^{\scriptscriptstyle SSGR}$ 
(mass dimension $1$ in 4d) corresponds to the mass parameter 
of the {\sc sbh}. (Note that $C$ has mass dimension $-2$ in 4d, and, 
hence, the need for the $\lambda^{-3}$ factor is evident.)


\section{ADM decompositions and kinematics for Zweibeine}


Let us now provide a more detailed description of 
the boundary structure $\partial {\cal M}$ associated with 
our {\em spatially bounded} spacetime region 
$\cal M$. The region $\cal M$ consists of a collection
of one dimensional spacelike slices $\Sigma$. The letter $\Sigma$
denotes both the foliation of $\cal M$ into spacelike slices and a
generic leaf of the foliation. The initial
spacelike slice is $\Sigma'$ (the curve of constant coordinate
time $t = t'$), and,
likewise, the final spacelike slice is $\Sigma''$ (the curve
of constant coordinate time $t = t''$).
On spacetime $\cal M$ we have coordinates $(t, r)$,
and a generic spacetime point is $B(t,r)$. 
Every constant-$t$ slice $\Sigma$
has two boundary points $B_{i}$ (at $r = r_{i}$) and $B_{o}$
(at $r = r_{o}$). Assume that along $\Sigma$ the coordinate $r$
increases monotonically from $B_{i}$ to $B_{o}$. We represent the
timelike history $B_{i}(t) \equiv B(t, r_{i})$ by $\bar{\cal T}_{i}$
and refer to it
as the {\em inner boundary}. Likewise, we represent the timelike
history $B_{o}(t) \equiv B(t, r_{o})$ by $\bar{\cal T}_{o}$ and refer
to it as the {\em outer boundary}.\footnote{Were we to adopt 
``thermodynamic boundary conditions''\cite{BMWY,Louko_Whiting,Lau}
for the black-hole solutions we consider, we would 
``seal" the inner boundary. In other words, the time development 
at the inner boundary would be arrested,
and the point $B_{i}$ would correspond to a bifurcation 
point in a Kruskal-like diagram. Although we do not consider
such boundary conditions in this work, we make mention of them,
since we do go some of the way towards setting them up.}
We denote the {\em corner} points of our
spacetime as follows: $B'_{i} \equiv B(t',r_{i})$, $B'_{o} \equiv
B(t',r_{o})$, $B''_{i} \equiv B(t'',r_{i})$, and $B''_{o} \equiv
B(t'',r_{o})$.

We can also consider a {\em radial} foliation of $\cal M$
by a family of $1$-dimensional timelike slices which extend
from $\bar{\cal T}_{i}$ outward to $\bar{\cal T}_{o}$ (here the
${\cal T}$'s stand for ``time'' and they are ``barred'' for 
reasons which become clear)\cite{GHayward_Wong}. 
These are constant-$r$ curves in ${\cal M}$.
Like before, we loosely use the letter $\bar{\cal T}$ both to denote
the radial foliation and a generic leaf of this foliation. 
Abusing the notation a bit, we also often let the symbol 
$\bar{\cal T}$ denote the total timelike boundary 
$\bar{\cal T}_{i} \bigcup \bar{\cal T}_{o}$. 
However, when the symbol $\bar{\cal T}$ has this meaning, it 
always appears in the phrase ``the $\bar{\cal T}$ boundary.''
Fig.~1 depicts the geometry associated 
with the foliations of our spacetime patch ${\cal M}$.

\subsection{Metric decompositions}
The metric on a generic
$\Sigma$ slice is $\Lambda^{2}$, and the metric on the
$\bar{\cal T}$ boundary is denoted by
$- \bar{N}^{2}$. In terms of the $\Sigma$ foliation, the
metric may be written in {\sc adm} form\cite{ADM}
\begin{equation}
g_{\mu\nu}{\rm d}x^{\mu} {\rm d}x^{\nu}
= - N^{2} {\rm d} t^{2} + \Lambda^{2}\left({\rm d}r + N^{r}
{\rm d}t\right)^{2}\, ,\label{ADM}
\end{equation}
with $N$ and $N^{r}$ denoting the familiar {\em lapse}
and {\em shift}. The vector field
\begin{equation}
u^{\mu} \partial/\partial x^{\mu} 
= N^{-1}\left(\partial/\partial t 
- N^{r} \partial/\partial r \right)
\end{equation}
is the unit, timelike, future-pointing normal to the 
$\Sigma$ foliation. 

In terms of the $\bar{\cal T}$ foliation, the $\cal M$ metric 
takes the form
\begin{equation}
g_{\mu\nu}{\rm d}x^{\mu} {\rm d}x^{\nu}
=  \bar{\Lambda}^{2} {\rm d} r^{2}
- \bar{N}^{2}\left({\rm d}t +
\bar{\Lambda}^{t}{\rm d}r\right)^{2}\, ,
\end{equation}
where $\bar{\Lambda}$ and 
$\bar{\Lambda}^{t}$ are the {\em radial
lapse} and {\em radial shift}. 
The unit, spacelike, $\bar{\cal T}$-foliation
normal is
\begin{equation}
\bar{n}^{\mu} \partial/\partial x^{\mu} = 
\bar{\Lambda}^{-1}\left(\partial/\partial r
- \bar{\Lambda}^{t}\partial/\partial t\right)\, .
\end{equation}
On the outer boundary $\bar{\cal T}_{o}$ the outward normal is
$\bar{n}^{\mu}$, while on the inner boundary
$\bar{\cal T}_{i}$ the outward normal is  
$- \bar{n}^{\mu}$. 

By equating the coefficients of the above forms of $g_{\mu\nu}$, we obtain
the following relations between the ``barred" and ``unbarred" variables:
\begin{eqnarray}
\bar{N} & = & N/\gamma \label{transformation}
\eqnum{\ref{transformation}a} \\
\bar{\Lambda} & = & \gamma \Lambda
\eqnum{\ref{transformation}b}\\
\Lambda N^{r}/N & = & - \bar{N}
\bar{\Lambda}^{t}/\bar{\Lambda}\, .
\eqnum{\ref{transformation}c}
\addtocounter{equation}{1}
\end{eqnarray}
Here $\gamma := (1 - v^{2})^{-1/2}$ is the
local relativistic factor
associated with the velocity $v := \Lambda N^{r}/ N
= - \bar{N} \bar{\Lambda}^{t}/\bar{\Lambda} 
=: - \bar{v}$\cite{GHayward_Wong,Lau,Lau2}. 
The timelike normal 
associated with the foliations 
$B_{i}(t)$ and $B_{o}(t)$ of the boundary slices
$\bar{\cal T}_{i}$ and $\bar{\cal T}_{o}$ is 
$\bar{u}^{\mu} \partial /\partial x^{\mu} 
= \bar{N}^{-1} \partial/\partial t$. Note
that on the $\bar{\cal T}$ boundary the vector fields 
$u^{\mu}$ and $\bar{u}^{\mu}$ need not
coincide. Also, fixation of the $t$ coordinate gives a
collection of points $B(r)$ which foliates the slice $\Sigma$.
The normal associated with this foliation of $\Sigma$ is
$n^{\mu} \partial/\partial x^{\mu} 
= \Lambda^{-1}\partial /\partial r $. 
At the inner boundary $- n^{\mu}$ is the outward-pointing 
normal of $B_{i}$ as embedded in
$\Sigma$, while at the outer boundary $n^{\mu}$ is the
outward-pointing normal of $B_{o}$ as embedded in $\Sigma$.
On the inner and outer boundaries $n^{\mu}$ and 
$\bar{n}^{\mu}$ need not coincide.
It is easy to verify that we have
\begin{eqnarray}
\bar{u}^{\mu} & = & \gamma u^{\mu} + v \gamma n^{\mu}
\label{boundaryboost} \eqnum{\ref{boundaryboost}a} \\
\bar{n}^{\mu} & = & \gamma n^{\mu} + v \gamma u^{\mu}
\eqnum{\ref{boundaryboost}b}
\addtocounter{equation}{1}
\end{eqnarray}
as the point-wise boost relations between the ``barred''
and ``unbarred'' frames.

With the 1-forms $u = u_{\mu} {\rm d}x^{\mu}$ and 
$n = n_{\mu}{\rm d}x^{\mu}$ we write down a general null 
spacetime zweibein, namely,
\begin{eqnarray}
  e^{+}
  & = & \sigL e^{\rho} 
  \sqrt{\textstyle \frac{1}{2}} 
  (- u + n) =
  \sigL e^{\rho} 
  \sqrt{\textstyle \frac{1}{2}} 
  [(N + \Lambda N^{r}) {\rm d}t 
  + \Lambda {\rm d}r]
\label{nullframe} \eqnum{\ref{nullframe}a} \\
  e^{-} & = & 
  \sigL e^{- \rho} 
  \sqrt{\textstyle \frac{1}{2}} 
  (- u - n) 
  = \sigL e^{-\rho} 
  \sqrt{\textstyle \frac{1}{2}} 
  [(N - \Lambda N^{r})
   {\rm d} t - \Lambda {\rm d}r]
{\,} ,
\eqnum{\ref{nullframe}b}
\addtocounter{equation}{1}
\end{eqnarray}
where $\sigL = \pm 1$ (``$L$'' for ``Lorentz'') and $\rho$ 
(an arbitrary point-wise boost factor) ensure that the zweibein 
is completely general. In terms of the 1-forms $\bar{u} 
= \bar{u}_{\mu} {\rm d}x^{\mu}$ and 
$\bar{n} = \bar{n}_{\mu}{\rm d}x^{\mu}$ our general zweibein is
given by
\begin{eqnarray}
  e^{+} & = & \sigL e^{\bar{\rho}} \sqrt{\textstyle \frac{1}{2}}
  (- \bar{u} + \bar{n})
= \sigL e^{\bar{\rho}} \sqrt{\textstyle \frac{1}{2}} 
  [\bar{N} {\rm d} t + 
  (\bar{N}\bar{\Lambda}^{t} + \bar{\Lambda}) {\rm d} r]
\label{nullframe2} \eqnum{\ref{nullframe2}a} \\
  e^{-} & = &  \sigL e^{- \bar{\rho}} 
\sqrt{\textstyle \frac{1}{2}}(-\bar{u} - \bar{n})
= \sigL e^{-\bar{\rho}} \sqrt{\textstyle \frac{1}{2}} [\bar{N}
   {\rm d} t + (\bar{N}\bar{\Lambda}^{t} - \bar{\Lambda})
   {\rm d} r]
{\,} ,
\eqnum{\ref{nullframe2}b}
\addtocounter{equation}{1}
\end{eqnarray}
with $\bar{\rho} = \rho + \eta$. The parameter 
$\eta = \frac{1}{2}\log|(1+v)/(1-v)|$ is associated with the local
boost between $u$ and $\bar{u}$. By construction, $\rho$, $\bar{\rho}$,
and $\eta$ are everywhere finite and well-defined.

\subsection{Killing vector}

For the models we study with the {\sc ecm} action, (at least in 
vacuo) the corresponding solution $g_{\mu\nu}$ will possess a 
Killing field $k^{\mu}$; and we shall be quite interested in 
the preferred foliation of spacetime which is associated with 
$k^{\mu}$. We use $\hat{\Sigma}$ both to denote this foliation 
and to represent a generic leaf thereof. Note that this 
foliation may have certain pathologies within the finite patch 
${\cal M}$ under examination. 
As experience with the Schwarzschild geometry suggests, $k^{\mu}$ 
need not be a timelike vector everywhere on our spacetime region. 
In general, a maximal extension of the solution $g_{\mu\nu}$ will 
include static and dynamical regions separated by horizons. 
{\em A priori} we want to
allow for all possibilities: that our spacetime patch ${\cal M}$
lies entirely within a dynamical region, that it lies entirely within
a static region, and that it covers portions of both static and
dynamical regions. Later on in $\S$ VI we shall find it 
necessary to assume that both of the timelike boundary elements,
$\bar{\cal T}_{i}$ and $\bar{\cal T}_{o}$, lie entirely within static
regions.

For vacuum, an equation of motion [namely, (\ref{eom}c) given later]
implies that $X^{\pm} = \pm e_{\mp}[X]$. Now, in fact, the 
vector fields $\sigL e_{\pm}{}^{\mu}$ are by construction everywhere 
future-pointing in our formalism. Hence, with the Penrose diagram for 
the {\sc sbh} and the above formulae as guides, we set $X^{\pm} = 
-\sigma_{\pm} \sigL |X^{\pm}|$ (defining $\sigma_{\pm}$) and introduce 
spacetime regions of type I, II, III, and IV as follows: 
\begin{eqnarray}
{\rm I}, & \hspace{1cm} & \sigplus = 1 = \sigminus \nonumber \\
{\rm II}, & &  \sigplus = 1 = - \sigminus \nonumber \\ 
{\rm III}, & & \sigplus = -1 = \sigminus \nonumber \\ 
{\rm IV}, & & \sigplus = -1 = -\sigminus
{\,} . \nonumber
\end{eqnarray}
Here the {\sc sbh} has been used only to elucidate the appropriate 
choices. Analogous ones are made for the (possibly quite 
complex) Penrose diagram corresponding to a given general {\sc ecm}. 
Let $\hat{u}^{\mu}$ represent the normalized Killing direction, 
i.~e.~
\begin{equation}
\hat{u}^{\mu} = |k_{\lambda} k^{\lambda} |^{-1/2} 
k^{\mu}
{\,} ,
\end{equation}
and pick its orthogonal partner $\hat{n}^{\mu}$ such that in a 
static region the pair 
$(\sigplus\hat{u}{}^{\mu}, \sigminus\hat{n}{}^{\mu})$ 
has the same orientation as $(u^{\mu}, n^{\mu})$. A static 
region is determined by $\sigK := - sign(k_{\mu}k^{\mu}) = 1$ 
(``$K$''for ``Killing''). Notice that $\sigplus\sigminus = \sigK$.
For a static region $\sigplus \hat{u}^{\mu}$ is 
the unit future-pointing 
$2$-velocity of Eulerian observers 
which ride along the orbits of the isometry and are 
instantaneously at rest in the $\hat{\Sigma}$ slices. 
However, in a dynamical region ($\sigK = -1$) we must exchange 
the roles of $\hat{u}^{\mu}$ and $\hat{n}^{\mu}$. In a 
dynamical region it is the pair  
$(\sigminus\hat{n}{}^{\mu}, \sigplus \hat{u}{}^{\mu})$ which has the
same orientation as $(u^{\mu}, n^{\mu})$.
Therefore, we find
\begin{eqnarray}
  \hat{u}^{\mu} & = & {\textstyle \frac{1}{2}} 
   (\sigplus + \sigminus)
   (\hat{\gamma} u^{\mu} + \hat{v} \hat{\gamma} n^{\mu})  
+ {\textstyle \frac{1}{2}}(\sigplus - \sigminus) 
  (\hat{\gamma} n^{\mu} + \hat{v}\hat{\gamma} u^{\mu}) 
\label{killingboost} \eqnum{\ref{killingboost}a} \\
  \hat{n}^{\mu} & = & {\textstyle \frac{1}{2}} (\sigminus + \sigplus)
   (\hat{\gamma} n^{\mu} + \hat{v} \hat{\gamma} u^{\mu})  
+ {\textstyle \frac{1}{2}}(\sigminus - \sigplus) 
  (\hat{\gamma} u^{\mu} + \hat{v}\hat{\gamma} n^{\mu}) 
\eqnum{\ref{killingboost}b}
{\,} 
\addtocounter{equation}{1}
\end{eqnarray}
as the boost relations between the $\Sigma$ Eulerian observers 
and the $\hat{\Sigma}$
Eulerian observers. Here, like before, $\hat{v}$ is a local boost 
velocity and $\hat{\gamma} = (1 - \hat{v}^{2})^{-1/2}$ is the associated
relativistic factor. Note that our construction breaks down on a 
horizon, characterized by $k_{\mu} k^{\mu} = 0$. Therefore,
we may write the arbitrary null zweibein as 
\begin{eqnarray}
  e^{+} & = & \sigL e^{(\rho + \hat{\eta})}\sqrt{\textstyle \frac{1}{2}}
  (- \sigminus \hat{u} + \sigplus \hat{n})
\label{tildenullzb}
\eqnum{\ref{tildenullzb}a}
\nonumber \\
e^{-}  & = &  
\sigL e^{-(\rho + \hat{\eta})} \sqrt{\textstyle \frac{1}{2}}
  (- \sigplus\hat{u} - \sigminus \hat{n}) 
{\,} ,
\eqnum{\ref{tildenullzb}b}
\end{eqnarray}
where the parameter $\hat{\eta} 
= \frac{1}{2} \log|(1 + \hat{v})/(1 - \hat{v})|$,
very important for our purposes, describes the local 
boost between the $\Sigma$ and $\hat{\Sigma}$ observers. 
We note the $\hat{\eta}$ is everywhere well-behaved and finite {\em except}
on a horizon where it is $\pm \infty$. For a Kruskal-like diagram, the 
value of $\hat{\eta}$ at the bifurcation point has a 
direction-dependent limit. Notice that the sign factors 
$\sigma_{\pm}$ 
present in (\ref{tildenullzb}) ensures that the expression
\begin{equation}
g := - e^{+} \otimes e^{-} - e^{-} \otimes e^{+}
  = \sigK(- \hat{u} \otimes \hat{u}
+ \hat{n} \otimes \hat{n})
\end{equation}
for the spacetime metric has the correct signature in all regions.
So far we have a three-fold {\em non}-degeneracy in our 
formalism. We have considered three distinct spacetime frames,
corresponding to $\Sigma$, $\bar{\Sigma}$, and $\hat{\Sigma}$.
Of course, this non-degeneracy carries over to the physical or 
``tilde'' geometry as well.


\section{Equations of motion and conservation law}


Taking the simple case when $\alpha$ is a constant (which, 
as we show later, results in no loss of generality for our 
purposes), we find the following {\sc eom}s corresponding to the 
{\sc ecm} action (\ref{PSMaction}):\cite{Kummer_Widerin}
\begin{eqnarray}
  {\rm d} X^{\pm} \pm \omega X^{\pm} & = & 
  \pm e^{\pm} V + A^{(m)\pm}
\label{eom} \eqnum{\ref{eom}a} \\
  {\rm d}e^{\pm} \pm \omega \wedge e^{\pm} & = & 
- \alpha X^{\pm} e^{+} 
  \wedge e^{-} 
\eqnum{\ref{eom}b} \\
  {\rm d}X & = &  
  X^{+} e^{-} 
- X^{-} e^{+} + A^{(m)}
\eqnum{\ref{eom}c} \\
  {\rm d}\omega & = & 
- \lambda^{2} ({\rm d}V_{0}/{\rm d}X)
  e^{+} \wedge e^{-} + B^{(m)}
{\,} ,
\eqnum{\ref{eom}d}
\addtocounter{equation}{1}
\end{eqnarray}
where one also has matter {\sc eom}s obtained by varying the matter fields 
in $L^{(m)}$. The terms $A^{(m)\pm}$, $A^{(m)}$, and $B^{(m)}$ are 
produced by varying the matter action $L^{(m)}$ with respect to 
$e^{\pm}$, $\omega$, and $X$, respectively. Notice that $B^{(m)}$ is 
non-zero whenever the matter contribution to the action depends 
on $X$, as is the case for {\sc ssgr}. Minimally coupled scalar 
fields and fermions in 2d have $A^{(m)} = 0$. Moreover, for
such a scalar field $S$, the the matter contributions in (\ref{eom}a) 
are $A^{(m)\pm} = \mp [{}^{*}(e^{\pm} \wedge {\rm d}S)] e^{\mp}$ 
(note that the quantity enclosed by square brackets is a scalar). 
For our purposes it suffices only to note that the terms
$A^{(m)\pm}$, $A^{(m)}$, and $B^{(m)}$ may be present, thereby 
providing contributions to $C^{(m)}$ in (\ref{conservation2}). As 
described in the introduction, starting with (\ref{eom}a) and 
(\ref{eom}c), one obtains (\ref{conservation}), from which immediately 
follows the ``absolute'' conservation law (\ref{conservation2}). 
This derivation shows that it is the quantity
\begin{equation}
C = X^{+} X^{-} \exp(\alpha X) + W
{\,} ,
\label{geometryC}
\end{equation}
and only this quantity, which is affected by the influx of 
matter\cite{Kummer_Widerin}. 
Regarding the presence of matter, for the following 
we only need this piece of information from the full {\sc eom}s.

The most general vacuum 
solution\cite{Schaller_Strobl,Kloesch_Strobl,Kummer_Widerin,Kummer_Schwarz} 
to the {\sc eom}s (\ref{eom}) is readily obtained 
\begin{eqnarray}
 e^{-} & = & - \exp (\alpha X) X^{-} {\rm d}f 
\label{PSMsolution}  \eqnum{\ref{PSMsolution}a} \\
  e^{+} & = & - {\rm d} X/ X^{-}
- X^{+} \exp (\alpha X) {\rm d} f
\eqnum{\ref{PSMsolution}b} \\
  \omega  & = & {\rm d}X^{-}/X^{-} 
- \exp (\alpha X)V {\rm d}f
\eqnum{\ref{PSMsolution}c}
\addtocounter{equation}{1}
\end{eqnarray} 
in terms of arbitrary functions $f$, $X$, 
$X^{\pm}$ (see Ref.~\cite{Kummer_Widerin} for a simple derivation). 
Here $X^{+}$ must be expressed in terms of $C$ by using 
(\ref{geometryC}). The form (\ref{PSMsolution}) of the solution 
is valid for $X^{-} \neq 0$. A similar form with 
$X^{-} \leftrightarrow X^{+}$ allows for the description of
$X^{+} \neq 0$ patches (possibly including points where 
$X^{-} = 0$).
The resulting line-element
\begin{equation}
({\rm d}s)^{2} = - 2 \exp (\alpha X)
[(C - W){\rm d} f^{2} + {\rm d}f {\rm d}X]
\label{PSMline-element}
\end{equation}
is expressed in what are essentially Eddington-Finkelstein
coordinates. The quantity
\begin{equation}
k^{\lambda} k_{\lambda} 
= - 2 \exp(\alpha X)( C - W ) 
\label{Killingnorm}
\end{equation}
coincides with the norm of the Killing vector field
$k^{\mu}$ [$= (\partial/\partial f)^{\mu}$ in the chosen
coordinates], and the presence and type of horizon associated with
the line-element (\ref{PSMline-element}) can be simply read off
from of the zeros in (\ref{Killingnorm}). Notice that
$C - W = \sigK | C - W |$ (cf.~$\S$ II.B). 

Consider again the Schwarzschild example (vacuum {\sc ssgr}), 
where we have set $C^{\scriptscriptstyle SSGR} = - 
\lambda^{3} M_{\scriptscriptstyle S}$ and $X = 
{\textstyle \frac{1}{2}}\lambda^{2} R^{2}$. From 
(\ref{SSGRchoices}b) and the definition of $W$, we find 
$W^{\scriptscriptstyle SSGR} 
= - {\textstyle \frac{1}{2}}\lambda^{3} R$. 
Therefore, in the vacuum {\sc ssgr} case (\ref{PSMline-element}) 
takes the form
\begin{equation}
({\rm d}s)^{2} = - \lambda R
[ (1 - 2M_{_{S}}/R) {\rm d}u^{2} + 2 {\rm d}u {\rm d}R]
\end{equation}
in terms of the retarded-time coordinate $u = \lambda f$. 
Upon our conformal rescaling $({\rm d}s)^{2} = \exp (2\varphi)
({\rm d}\tilde{s})^{2}$ with $\exp (2 
\varphi^{\scriptscriptstyle SSGR}) = \lambda R$
from (\ref{SSGRchoices}c), we obtain $({\rm d}\tilde{s})^{2}$,  
the time-radial piece of the {\sc sbh} line-element in 
outgoing EF coordinates\cite{MTW}. We make the important 
observation that for {\sc ssgr} the conformal factor is 
$\exp(2\varphi^{\scriptscriptstyle SSGR}) = 
2\lambda^{-2} |W^{\scriptscriptstyle SSGR}|$. One recovers
the {\sc dbh} line-element in a similar 
fashion.\cite{Kloesch_Strobl,Kummer_Widerin}

By comparing our expressions for $e^{\pm}$ in terms of the
{\sc adm} variables with the general solution (\ref{PSMsolution}), we 
obtain the following expressions for the $\Sigma$ metric, lapse, 
and shift
\begin{eqnarray}
\Lambda^{2} & = & 
            - 2 \exp(\alpha X) 
              f'\left[X' + (C - W) f'\right]
\eqnum{\ref{lapseshiftLambda}a} \\
  N & = &      \exp(\alpha X)\Lambda^{-1}
               \left[\dot{f} X' - \dot{X}f'\right]
\label{lapseshiftLambda}
\eqnum{\ref{lapseshiftLambda}b} \\
  N^{r} & = & - \exp (\alpha X)\Lambda^{-2}\left[\dot{X}f' 
+ \dot{f}X' + 2 (C - W)\dot{f}f'\right]
{\,} .
\eqnum{\ref{lapseshiftLambda}c} 
\addtocounter{equation}{1}
\end{eqnarray}
Let us now verify that $N$ and $\Lambda^{2}$ are everywhere 
positive, which should follow by our assumption that the 
{\em everywhere-spacelike} slices of the $\Sigma$ foliation
{\em everywhere advance} into the future. To this end, 
let $f = \tau - X^{*}$, where $X^{*}$ is a tortoise-type 
coordinate\cite{MTW} obeying $\partial X^{*}/\partial X
= \frac{1}{2}(C - W)^{-1}$; and consider the coordinates
$(\tau, X)$. As experience with the Schwarzschild geometry
suggests, in a static region ($\sigK = 1$) we know that 
either the coordinate set $(\tau,X)$ or $(-\tau, -X)$ will have 
the same orientation as the set $(t,r)$ of coordinates associated
with the $\Sigma$ foliation. In a dynamical region either the
set $(-X, \tau)$ or $(X, -\tau)$ will have the same orientation
as $(t,r)$. Using these coordinates along with the sign of
the Killing norm (\ref{Killingnorm}), one may straightforwardly
verify that both $N$ and $\Lambda^{2}$ are everywhere positive
in all regions covered by our spacetime patch ${\cal M}$.


\section{Dilaton Models with Minkowski Ground State}


Kl\"{o}sch and Strobl have carried out a comprehensive global 
analysis of all vacuum 2d gravity models based upon the general
solution (\ref{PSMsolution})\cite{Kloesch_Strobl}. Their
results, as well as previous ones obtained from the global
analysis of ``$R^{2} + T^{2}$ gravity''\cite{Katanaev_Volovich},
clearly demonstrate that, in general, the singularity structures 
for such models exhibit little similarity with the 2d singularity 
structures familiar from {\sc ssgr}, such as the {\sc sbh} and {\sc rnbh}. 
In particular, a flat (or at
least de Sitter) ``ground state'' for certain values of $C$ is
not realized. Moreover, even asymptotically flat (or 
asymptotically de Sitter) solutions are difficult to realize.
However, as mentioned, by a (local) re-definition of fields it 
has been shown that {\sc ecm} models [even those with torsion, 
$\alpha \neq 0$ in (\ref{PSMpotential})] are equivalent to 
torsion-free {\sc gdt}s.\cite{KKL1} In this correspondence
between {\sc ecm}s and {\sc gdt}s, the resulting change in global properties
gives rise to the appearance of asymptotically flat solutions.

For some time, the {\sc dbh} from 
{\footnotesize 2d}{\sc dg} \cite{Sengupta_Wadia,Witten,CGHS} 
has attracted
special interest because several of its features, namely,
asymptotic flatness and a Minkowski ground state for $C = 0$,
mimic those associated with the {\sc sbh}. Moreover, the {\sc dbh} 
in the presence of matter is also an exactly solvable model. 
Recently, however, it has 
been shown that the original {\sc dbh} resembles the real {\sc sbh} 
rather superficially, in that it possesses complete null 
geodesics\cite{KKL2}. In Ref.~\cite{KKL2}, a family of models 
has been considered which are characterized by one bifurcate 
Killing horizon as well as incomplete non-null and 
\underline{null} geodesics. In other words, each model in 
the family closely resembles the {\sc sbh} (which is, in fact, a 
member of the class itself) and possesses a ``genuine''
{\sc sbh} Penrose diagram. We would like to point out that these
models are among the {\sc gdt}s we study here.

\subsection{Physical dilaton models}
Motivated by the fact that theories with torsion are (locally) 
equivalent to (globally better-behaved) generalized dilaton
models, we consider only the torsion-less case here
[${\alpha} = 0$ in (\ref{PSMpotential})]. As we shall explicitly
demonstrate later in $\S$ VI.A, this assumption results in no 
loss of generality at the level of {\sc gdt}s (i.~e.~after 
conformal transformation).
If we demand that the line-element for $C = 0$ reduces to the one 
for flat spacetime, then we find an (essentially) unique conformal 
transformation of (\ref{PSMsolution}). Namely, the one determined
by
\begin{equation}
2\varphi = \log (2\lambda^{-2} |W|)
\label{special_conformal_factor}
\end{equation}
(the factor of 2 under the square root is chosen to 
have agreement with the standard conventions for {\sc ssgr})
which leads to the following rescaled metric:
\begin{equation}
({\rm d}\tilde{s})^{2} = 
- [(\sigw + C|W|^{-1})\lambda^{2}{\rm d}f^{2} + 
2{\rm d}f {\rm d}\tilde{X}]
{\,} .
\label{flat-line-element}
\end{equation}
Here we have set $W = - \sigw |W|$ 
and defined $\tilde{X}$ by
\begin{equation}
{\rm d}\tilde{X}/{\rm d}X = 
{\textstyle \frac{1}{2}}\lambda^{2}|W|^{-1} 
{\,} .
\end{equation}
One obtains an asymptotically flat line-element for models with
$\sigw = 1$ (and, hence, negative $W$) and $\lim_{X\to\infty} 
|W| = \infty$. Henceforth, we shall be most interested in the
most general class of such models. The models in this general class 
are the ones that we refer to as ``physical'' (cf.~the 
discussion in $\S$ I.A and I.C). 
In terms of the new coordinate $\tilde{X}$, we rewrite the {\sc gdt} 
action (\ref{hatdilatonaction}) in true ``dilaton form,''
\begin{eqnarray}
  \tilde{L}|^{\rm dil} & = &
  {\textstyle \frac{1}{2}}
  \kappa \int_{\cal M} 
  {\rm d}^{2}x \sqrt{-\tilde{g}}
  \left[X \tilde{\cal R} 
- \left({\rm d}^{2}X/{\rm d}\tilde{X}{}^{2}\right) 
  \left(\tilde{g}^{\mu\nu}
   \partial_{\mu} \tilde{X} \partial_{\nu} \tilde{X}
-  \lambda^{2}\right)\right]
\nonumber \\
& &
+ \kappa \oint_{\partial{\cal M}}{\rm d}x 
  \sqrt{|\tilde{g}^{1}|} 
  X \tilde{\epsilon}^{\mu\nu} 
  \tilde{\omega}_{\mu} \tilde{\sf n}_{\nu} + L^{(m)}
{\,} .
\label{hatdilatonaction2}
\end{eqnarray}
In this action one should consider $X = X(\tilde{X})$. 
As mentioned, unless we employ a suitable gauge choice for
the zweibein at the boundary $\partial {\cal M}$, the
action (\ref{hatdilatonaction2}) cannot be viewed solely as
a metric action. With strict attention paid to this point, 
we examine the canonical form of this action in $\S$ VI.A.

Using our earlier results (\ref{lapseshiftLambda}) we may obtain
expressions for the rescaled $\Sigma$ metric 
$\tilde{\Lambda}^{2} = {\textstyle \frac{1}{2}}\lambda^{2}|W|^{-1}
\Lambda^{2}$, lapse $\tilde{N} = 
\sqrt{\textstyle \frac{1}{2}}\lambda|W|^{-1/2} N$, and
shift $\tilde{N}^{r} = N^{r}$. Indeed,
a short calculation yields the following set:
\begin{eqnarray}
  \tilde{\Lambda}^{2} & = & 
- f'\left[2\tilde{X}' 
+ (1 + C|W|^{-1}) \lambda^{2} f'\right]
\eqnum{\ref{tildelapseshiftLambda}a} \\
\tilde{N} & = & \tilde{\Lambda}^{-1} 
                \left[\dot{f} \tilde{X}' - \dot{\tilde{X}} f'
                 \right]
\label{tildelapseshiftLambda}
\eqnum{\ref{tildelapseshiftLambda}b} \\
  N^{r} & = & - \tilde{\Lambda}^{-2}
                \left[
                \dot{\tilde{X}}f' 
              + \dot{f}\tilde{X}' + (1 
+ C|W|^{-1})\lambda^{2} \dot{f}f'\right]
{\,} .
\eqnum{\ref{tildelapseshiftLambda}b} 
\addtocounter{equation}{1}
\end{eqnarray}
Now, since conformal transformations preserve casual structure,
we need not check the signs of $\tilde{N}$ and $\tilde{\Lambda}^{2}$
as we did with the unphysical lapse and $\Sigma$ metric. We have
written these equations in order to make the following derivation.
Setting $C = 0$ we obtain a flat line-element. Therefore, we use
coordinate freedom such that $(t, r)$ produce a manifestly flat 
line-element, $({\rm d}\tilde{s}_{0})^{2} = - {\rm d}t^{2} 
+ {\rm d}r^{2}$. This implies that $\tilde{N}_{0} = 
\tilde{\Lambda}_{0}^{2} = 1$ and 
$\tilde{N}_{0}^{r} = 0$. Enforcing these conditions on the set
(\ref{tildelapseshiftLambda}), we obtain 
\begin{equation}
\tilde{X}_{0} = \lambda r
{\,} .
\label{tildeXnaught}
\end{equation}
Therefore, the groundstates of the {\sc gdt} models we consider are
in fact ``linear-dilaton vacua'' as in the standard 
{\footnotesize 2d}{\sc dg}.\cite{Witten,CGHS} The existence of such a groundstate
plays a crucial role in our examination of asymptotic energy
found in $\S$ VI.C. One also finds that $u_{0} := \lambda f_{0}
= t - \lambda^{-1}\tilde{X}_{0}$ as expected.

We should mention that the influx of particles into the 
groundstate with $C = 0$ changes 
the state to $C \neq 0$. Therefore, as in all strictly 2d models 
we encounter the well-known deficiency that there is no critical 
density below which an ingoing ``radial'' wave is simply reflected 
to an outgoing one (i.~e.~$C$ stays zero) so that no 
singularity is created. This would appear to be cured by the 
introduction of non-minimal couplings (e.~g.~involving an 
explicit $X$-dependence as in {\sc ssgr}). In {\sc ssgr} criticality has been 
observed in some analytic (albeit rather unphysical) 
solutions\cite{Brady}. 

\subsection{Universal mass in GDT}

We conclude this section by discussing the concept of 
``universal mass'' in {\sc gdt}. With our conventions the well-known 
Misner-Sharp definition\cite{Misner_Sharp} of gravitational mass 
for {\sc ssgr} reads as follows:
\begin{equation}
M_{\scriptscriptstyle MS} 
= {\textstyle \frac{1}{2}}
  R\left[1 - \tilde{g}^{\mu\nu} 
  (\partial_{\mu}R)(\partial_{\nu}R)\right]
{\,} ,
\end{equation}
again where $R$ is the areal radius. This is the accepted 
definition of ``quasilocal'' gravitational mass in {\sc ssgr}, and 
a wide array of quasilocal mass constructions from full 
4d general relativity yield this result in the 
special case of spherical 
symmetry.\cite{manypeople2} Viewed as a 
canonical expression (cf.~the discussion in $\S$ V, VI), 
the Misner-Sharp mass plays a fundamental role in Kucha\v{r}'s 
approach\cite{Kuchar} to the canonical reduction of (vacuum) 
{\sc ssgr}.\footnote{Kucha\v{r}'s approach is the {\sc sbh}
version of the canonical reduction we shall study in $\S$ VI.
A quite similar analysis was earlier carried out by Schaller and 
Strobl\cite{new_Strobl} in the context of the Katanaev-Volovich 
model\cite{Katanaev_Volovich}.}
Such a ``mass canonical variable'' had been earlier isolated 
and successfully used in the canonical reduction of {\sc ssgr} 
by Kastrup and Thiemann. Their work is based mostly, but not 
entirely, on the Ashtekar gravitational 
variables\cite{Kastrup_Thiemann}. Furthermore, 
Guven and \'{O} Murchadha's thorough 
investigation\cite{Guven_O_Murchadha} of 
gravitational collapse 
in {\sc ssgr} also makes use of this mass. 

In our formalism a generalization of the Misner-Sharp 
definition, a {\em universal mass function} valid for all 
dilaton theories, appears quite naturally. Namely, the 
covariant expression for the quantity $C$ (absolutely 
conserved in vacuo). Again, set $\alpha = 0$, which,
to repeat, amounts to no loss of generality 
at the level of {\sc gdt}. We seek a covariant 
version of this expression which is written in terms of the 
conformally rescaled geometry. A simple analysis of the based
on the line-element (\ref{flat-line-element}) readily yields 
the desired expression,
\begin{equation}
C = {\textstyle \frac{1}{4}}
    \lambda^{2} |W|^{-1} \tilde{g}^{\mu\nu}
    (\partial_{\mu} X)(\partial_{\nu} X) - |W|
{\,} .
\end{equation}
With the {\sc ssgr} choices (\ref{SSGRchoices}) one may easily verify
that $C^{\scriptscriptstyle SSGR} = - \lambda^{3} 
M_{\scriptscriptstyle MS}$, as expected. The covariant expression
for $C$ takes an especially nice form in terms of the dilaton
field $\tilde{X}$ introduced above, namely, 
\begin{equation}
C = |W| \left[\lambda^{-2} \tilde{g}^{\mu\nu} 
    (\partial_{\mu} \tilde{X})(\partial_{\nu} \tilde{X}) 
    - 1 \right]
\end{equation}
For {\footnotesize 2d}{\sc dg} this is Tada and Uehara's 
``local mass''\cite{Tada_Uehara}, which has also been examined 
by Hayward and earlier by Frolov.\cite{SHayward,Frolov} As 
a ``mass canonical variable'' the universal mass has also been 
consider by several authors in the 
context of {\sc gdt}\cite{Martinez_Kunstatter,Barvinsky_Kunstatter}.


\section{Canonical Action and  Quasilocal Energy-Momentum}


In this section we begin our examination of the canonical 
formulation of both {\sc ecm}s and {\sc gdt}s for 
the bounded spacetime 
region ${\cal M}$. First, we present our variational principle,
which is determined by adding certain further boundary terms to the 
{\sc ecm} action, and then write the action 
in canonical form.
In this section we also examine the notion of gravitational 
(quasilocal) energy and momentum for both {\sc ecm}s and {\sc gdt}s.
Again, the vehicle for this examination is the 
first-order approach.
 
\subsection{Boundary terms and variational principle}

As mentioned, we are chiefly interested in those physical 
geometries $({\cal M},\tilde{g}_{\mu\nu})$ introduced in
$\S$ IV.A and obtained via the conformal transformation 
discussed in $\S$ I.C. We have argued that the addition 
of a boundary term to 
the base {\sc ecm} action is necessary in order to preserve 
the variational principle under the conformal 
rescaling.\footnote{We shall show in the next section, that
from the canonical viewpoint the addition of a boundary term
to the base {\sc ecm} action is necessary in order to manifest the 
conformal rescaling as a canonical transformation.}
However, note that the naive boundary term 
(\ref{naive_surface_term}) introduced in $\S$ I.C  is in 
fact zweibein-gauge dependent, and this breaks the local
Lorentz invariance of the theory at the boundary. In this
section we shall write down an ``improved'' boundary term
which, just like the naive boundary term, ensures preservation 
of the variational principle under the conformal 
rescaling of $\S$ I.C. However, our improved boundary term is
zweibein-gauge independent (in a sense to be made precise below). 
Therefore, as we shall explicitly 
demonstrate in the canonical formalism, our improved boundary
term maintains complete zweibein-gauge freedom at the 
boundary.  Moreover, as we will also show in the next
section, our improved boundary term affords a rigorous 
correspondence between {\sc ecm}s ({\em zweibein} theories) and 
{\sc gdt}s ({\em metric} theories).

In order to write down the improved boundary term, we first 
collect a few definitions. Recall that we may write the 
{\em general}-gauge 
spacetime dual-zweibein as either (\ref{nullframe}) or
(\ref{nullframe2}), where the connection one-form associated with 
the arbitrary gauge is $\omega$. Define the {\em time gauge} by 
setting $\rho = 0$, $\sigL  = 1$, and denote the connection 
$1$-form for this special gauge by 
$\Omega = \omega + {\rm d}\rho$. Also, 
define the {\em radial gauge} by setting $\bar{\rho} = 0$, 
$\sigL = 1$, and denote 
the connection $1$-form for this special gauge by $\bar{\Omega} = 
\omega + {\rm d}\bar{\rho}$. Clearly, as stated before in $\S$ II, 
the parameter associated with the local boost between 
the time-gauge and radial gauge zweibeine is 
$\eta = \bar{\rho} - \rho$. Our improved boundary term is 
\begin{equation}
  L_{\partial {\cal M}} =
- \kappa \oint_{\partial {\cal M}} X\omega
+ \kappa \int^{\Sigma''}_{\Sigma'}\rho {\rm d}X
- \kappa \int_{\bar{\cal T}}\bar{\rho} {\rm d}X
{\,} ,
\label{improvedsurfaceterm}
\end{equation}
where in this expression one should realize that $\rho$ and
$\bar{\rho}$ are short-hand for 
$\frac{1}{2}\log (- e^{+}{}_{r}/e^{-}{}_{r})$ and
$\frac{1}{2} \log (e^{+}{}_{t}/e^{-}{}_{t})$, respectively.
Note that both 
$-e^{+}{}_{r}/e^{-}{}_{r}$ and $e^{+}{}_{t}/e^{-}{}_{t}$
are everywhere positive, so these logarithms make sense.
Moreover, as both $\rho$ and $\bar{\rho}$ are conformally
invariant, the extra two terms appearing on 
the right-hand side of (\ref{improvedsurfaceterm}) 
do not spoil the nice property that addition of 
(\ref{improvedsurfaceterm})
to the {\sc ecm} action (\ref{PSMaction}), like addition of 
the naive boundary term (\ref{naive_surface_term}),
ensures conformally preserved boundary conditions.
In (\ref{improvedsurfaceterm}) and below we adopt the 
following convention for integration:
\begin{equation}
 \oint_{\partial{\cal M}}  =  
  \int^{\Sigma''}_{\Sigma'} 
- \int_{\bar{\cal T}} + {\,} \Bigl. \Bigr|^{B''}_{B'}
\label{orientation}
\end{equation}
with
\begin{eqnarray}
  \int^{\Sigma''}_{\Sigma'} = 
  \int_{\Sigma''} 
- \int_{\Sigma'}
  &  & {\rm and,\,\, for\,\, example,}\,\, 
  \int_{\Sigma'} = \int^{r_{o}}_{r_{i}}\,\,
  {\rm at}\,\, t = t' 
\label{integration} \eqnum{\ref{integration}a} \\
  \int_{\bar{\cal T}} =
  \int_{\bar{\cal T}_{o}} 
- \int_{\bar{\cal T}_{i}}
 & & {\rm and,\,\,for\,\, example,} \,\, 
  \int_{\bar{\cal T}_{i}} = \int^{t''}_{t'}\,\,
  {\rm at}\,\, r = r_{i}
\eqnum{\ref{integration}b} \\
  \Bigl. {} \Bigr|^{B''}_{B'}&  = &  
  \Bigl. {} \Bigr|^{B_{o}''}_{B_{o}'} 
- \Bigl. {} \Bigr|^{B_{i}''}_{B_{i}'}
{\,} . 
\eqnum{\ref{integration}c}
\addtocounter{equation}{1}
\end{eqnarray}
[These conventions ensure ``clockwise'' integration over
the entire boundary $\partial {\cal M}$ for the integrals
and corner terms in (\ref{symbolic_traceK}) 
below (see Fig.~1).]
Because the corner points are a set of measure zero
in $\partial {\cal M}$, the term $|^{B''}_{B'}$ in 
(\ref{orientation}) is irrelevant unless the quantity
being integrated over the boundary has distributional 
support at the corners. In particular, there are {\em no}
corner-point contributions to 
$- \kappa \oint_{\partial{\cal M}}X\omega$ in
(\ref{improvedsurfaceterm}).

To gain insight into the geometric meaning of the boundary term
(\ref{improvedsurfaceterm}), on $\Sigma'$ and $\Sigma''$ 
replace $\omega$
with $\Omega - {\rm d}\rho$, and on the $\bar{\cal T}$ 
boundary replace $\omega$ with $\bar{\Omega} - {\rm d}
\bar{\rho}$. Next, integrate by parts to shift the ${\rm d}$'s
off of the boost parameters $\rho$ and $\bar{\rho}$ and onto
$X$'s, thereby
creating corner terms at $B''_{o}$, $B'_{o}$, $B''_{i}$, 
and $B'_{i}$. The result of this calculation is
\begin{equation}
  L_{\partial {\cal M}} = 
  \kappa \int^{\Sigma''}_{\Sigma'}{\rm d}r\Lambda X K
- \kappa \int_{\bar{\cal T}}{\rm d}t\bar{N} X \bar{\Theta} 
- \left. \kappa X \eta \right|^{B''}_{B'}
{\,},
\label{traceK}
\end{equation}
where, respectively, $K := -\Omega[n] 
= - \nabla_{\mu}u^{\mu}$ and $\bar{\Theta} 
:= - \bar{\Omega}[\bar{u}] 
= - \nabla_{\mu} \bar{n}{}^{\mu}$ are by definition the
mean curvatures associated with $\Sigma''$ (or $\Sigma'$)
as embedded in ${\cal M}$ and $\bar{\cal T}_{o}$ 
(or $\bar{\cal T}_{i}$) as
embedded in ${\cal M}$. As before, the variable $\Lambda$ is the 
square root of the induced $\Sigma$ one-metric $\Lambda^{2} 
= - 2e^{+}{}_{r} e^{-}{}_{r}$, and, likewise, $\bar{N}$ is the 
square root of minus the induced $\bar{\cal T}$ one-metric 
$\bar{N}^{2} = 2e^{+}{}_{t} e^{-}{}_{t}$. Therefore,
perhaps rather symbolically, we may write
\begin{equation}
  L_{\partial {\cal M}} =
 \kappa \oint_{\partial {\cal M}} {\rm d}x \sqrt{|g^{1}|}X
  {\cal K}
{\,} ,
\label{symbolic_traceK}
\end{equation}
with ${\cal K}$ denoting the 
mean curvature of the total boundary $\partial {\cal M}$
as embedded in ${\cal M}$ [in a generalized sense to the usual
notion of mean curvature from (pseudo)Riemannian geometry, as
$\nabla_{\mu}$ is the covariant derivative associated with a 
connection which may have torsion]. In other words, 
(\ref{improvedsurfaceterm}) is 
the ``$X$-averaged'' total mean curvature of the boundary as 
embedded in spacetime. The geometric importance of such a boundary 
term is well-known\cite{York,GHayward_Wong,Lau2,Brill_Hayward}. 
In particular, such a term is zweibein-gauge independent, 
i.~e.~depends solely on metric variables (and here possibly torsion).
Although the corner points constitute a set of measure zero in 
the integration of $X {\cal K}$ over the whole spacetime boundary, 
there are nevertheless finite contributions to 
the integral from these points as the $\partial {\cal M}$ 
normal changes discontinuously at these points, 
e.~g.~from $u^{\mu}$ to 
$\bar{n}^{\mu}$. Such corner terms are key in Brill and Hayward's 
treatment of the additivity of the 4d Einstein-gravity action 
under the composition of adjoining spacetime 
regions\cite{Brill_Hayward}.
We note that our boundary term 
(\ref{improvedsurfaceterm}) differs from the
type considered by Chan {\em et al}\cite{Chan_Creighton_Mann}
mainly in that we take the corner terms explicitly into account.

Adding the boundary term (\ref{improvedsurfaceterm}) to the base 
part of the {\sc ecm} action, we obtain
\begin{equation}
  L = \kappa\int_{\cal M}(- X_{a} De^{a} 
+ \omega \wedge {\rm d}X - V\epsilon)
+ \kappa \int^{\Sigma''}_{\Sigma'} \rho {\rm d}X 
- \kappa \int_{\bar{\cal T}} \bar{\rho} {\rm d}X 
{\,} 
\label{correctPSMaction}
\end{equation}
for the total action. (For simplicity, we ignore the matter
contribution $L^{(m)}$ to the action here. As long as the matter
is minimally coupled, its inclusion is straightforward.) By direct,
if tedious, calculation we find the following for the variation
of this action:
\begin{eqnarray}
  \delta L & = & [{\rm terms\,\,giving\,\,
  the\,\,{\sc eom}s\,\,(\ref{eom})}]
\nonumber \\ & & 
+ \int^{\Sigma''}_{\Sigma'} 
  {\rm d}r(P_{+} \delta e^{+}{}_{r}
+ P_{-} \delta e^{-}{}_{r} 
+ P_{X} \delta X) 
\nonumber \\ & &
+ \int_{\bar{\cal T}} {\rm d}t 
  (\bar{\Pi}_{+} \delta e^{+}{}_{t}
+ \bar{\Pi}_{-} \delta e^{-}{}_{t} 
+ \bar{\Pi}_{X} \delta X)
- \left.\kappa \eta \delta X \right|^{B''}_{B'}
{\,} ,
\label{variation}
\end{eqnarray}  
where we have defined the momenta\footnote{The momenta appearing in 
(\ref{momenta0}) can easily be rewritten in terms of components, 
for example, $P_{+} = \kappa(X^{-} + \frac{1}{2} X'/e^{+}{}_{r})$ and 
$P_{X} = -\kappa (\omega_{r} + \rho')$. In fact, to obtain 
the variation (\ref{variation}) of the action, it is easiest to
work with the canonical form
(\ref{canonicalaction1}) of the action, with the understanding 
that the momenta are short-hand for the aforementioned component
expressions.}
[cf.~the footnote just after Eq.(\ref{PSMaction}) for a 
description of the notation]
\begin{eqnarray}
  P_{\pm} & = & 
  \kappa 
  \left( X^{\mp} \mp n^{\mp}n[X]\right)
\label{momenta0} \eqnum{\ref{momenta0}a} \\
  P_{X} & = & 
- \kappa\Lambda\left( \omega[n] 
+ n[\rho]\right)
\eqnum{\ref{momenta0}b} \\
  \bar{\Pi}_{\pm} & = & 
- \kappa \left( X^{\mp} 
\mp \bar{u}^{\mp}\bar{u}[X]\right)
\eqnum{\ref{momenta0}c} \\  
  \bar{\Pi}_{X} & = & \kappa \bar{N} \left( \omega[\bar{u}] 
+ \bar{u}[\bar{\rho}]\right) {\,}.
\eqnum{\ref{momenta0}d}
\addtocounter{equation}{1} 
\end{eqnarray} 
Here and in what follows, we make extensive use of the notation
$n^{\pm} := n^{\mu} e^{\pm}{}_{\mu}$ (and similarly for 
$\bar{u}^{\pm}$, $u^{\pm}$, and $\bar{n}^{\pm}$).
Inspection of the variation (\ref{variation}) of the action 
shows that, with respect to a generic spacelike $\Sigma$ 
slice, $P_{\pm}$ and $P_{X}$ are the gravitational momenta
respectively conjugate to $e^{\pm}{}_{r}$ and $X$. Likewise, 
$\bar{\Pi}_{\pm}$ and $\bar{\Pi}_{X}$ are conjugate to 
$e^{\pm}{}_{t}$ and $X$ respectively, where now conjugacy is 
defined with respect to the $\bar{\cal T}$ boundary. At least
heuristically, $-\kappa\eta$ is the momentum conjugate to $X$ 
at the corners. Notice that the variational principle 
associated with the action (\ref{correctPSMaction}) features 
fixation of $X$ at the corners, in harmony with the fact that 
$X$ is also fixed on $\Sigma'$, $\Sigma''$, and the $\bar{\cal T}$
boundary. By construction, $-2n^{+} n^{-} = 1$, 
$2\bar{u}^{+} \bar{u}^{-} = 1$. Note also that the boost 
parameters above are in fact
$\rho = \frac{1}{2} \log (- n^{+}/n^{-})$ and $\bar{\rho} 
= \frac{1}{2}\log (\bar{u}^{+}/\bar{u}^{-})$.

The gravitational momenta $\{P_{\pm}, P_{X}\}$ are not the standard
ones usually associated with the canonical formulation of 
first-order 
{\sc ecm}s\cite{Kloesch_Strobl,Kummer_Widerin} 
(the standard {\sc ecm} canonical 
variables appear in the next section). 
Therefore, a few words are in order concerning their origin and 
geometric significance. First, $P_{X}$ is clearly the general-gauge 
expression for $- \kappa\Lambda\Omega[n]$. To gain insight into 
the geometric meanings of $P_{\pm}$, insert the identities 
$\delta e^{\pm}{}_{r} = 
n^{\pm}(\delta \Lambda \pm \Lambda \delta \rho)$
and
$\delta e^{\pm}{}_{t} = \bar{u}^{\pm} (\delta \bar{N} 
\pm \bar{N} \delta \bar{\rho})$ into the variation
(\ref{variation}) in order to find
\begin{eqnarray}
 (\delta L)_{\partial {\cal M}} & = & 
  \int^{\Sigma''}_{\Sigma'} 
  {\rm d}r(P_{\Lambda} \delta \Lambda
+ P_{\rho} \delta \rho + P_{X} \delta X) 
\nonumber \\ & &
+ \int_{\bar{\cal T}} {\rm d}t 
  (\bar{\Pi}_{\bar{N}} \delta \bar{N}
+ \bar{\Pi}_{\bar{\rho}} 
  \delta \bar{\rho} + \bar{\Pi}_{X} \delta X)
- \left.\kappa \eta \delta X \right|^{B''}_{B'}
{\,} ,
\label{alternatevariation}
\end{eqnarray}
where the alternative momenta are the following:
\begin{eqnarray}
  P_{\Lambda} & = & \kappa (X^{+} n^{-} + X^{-} n^{+})
\label{momenta1} \eqnum{\ref{momenta1}a} \\
  P_{\rho} & = & 
  \kappa \Lambda \left(n[X] 
- X^{+} n^{-} + X^{-} n^{+}\right)
\eqnum{\ref{momenta1}b} \\
  \bar{\Pi}_{\bar{N}} & = & 
  -\kappa (X^{+} \bar{u}^{-} + X^{-} \bar{u}^{+})
\eqnum{\ref{momenta1}c} \\
\bar{\Pi}_{\bar{\rho}} & = & 
- \kappa \bar{N}\left(\bar{u}[X] 
- X^{+} \bar{u}^{-} + X^{-} \bar{u}^{+}\right)
\eqnum{\ref{momenta1}d} {\,} .
\addtocounter{equation}{1}
\end{eqnarray}
With respect to a generic spacelike $\Sigma$ slice, 
the gravitational momentum $P_{\Lambda}$ is conjugate to
$\Lambda$. Likewise, with respect to the $\bar{\cal T}$ boundary,
the momentum $\bar{\Pi}_{\bar{N}}$ is conjugate to $\bar{N}$. 
As we shall see later, the expression 
for $- \bar{\Pi}_{\bar{N}}$ is closely related to the notion of 
quasilocal energy in the formalism. Notice that the {\sc eom} (\ref{eom}c) 
shows that both $P_{\rho}$ and $\bar{\Pi}_{\bar{\rho}}$ 
vanish on-shell. Therefore, $\rho$ and $\bar{\rho}$ need not be held 
fixed at the boundary in the variational principle associated with 
the action (\ref{correctPSMaction}), as the equation of motion (\ref{eom}c) 
ensures that both $P_{\rho}$ and $\bar{\Pi}_{\bar{\rho}}$ vanish for 
arbitrary
variations $\delta\rho$ and $\delta \bar{\rho}$ about a classical
solution. Hence, for our choice (\ref{correctPSMaction})
of action, the presence of the boundary 
$\partial {\cal M}$ does not break the zweibein-gauge freedom 
of the theory. In particular, this merely reflects the fact that $\rho$ 
does not represent a true dynamical degree of freedom. In terms
of the alternative canonical momenta (\ref{momenta1}) we find that
\begin{equation}
  P_{\pm} = - n^{\mp} (P_{\Lambda} \pm P_{\rho}/\Lambda)
{\, .}
\end{equation}
Therefore, modulo $P_{\rho}$ which vanishes on-shell, both $P_{\pm}$
are closely related to $P_{\Lambda}$. We may consider 
the canonical
pairs 
\begin{equation}
\{P_{\pm}, e^{\pm}{}_{r}; P_{X} , X\}
\label{Sigmavariables}
\end{equation} 
as $\Sigma$-adapted canonical variables ($\Sigma${\sc cv}s) which are 
nevertheless expressed in a general zweibein gauge. We have 
obtained these $\Sigma${\sc cv}s rather than the standard 
ones\cite{Schaller_Strobl,Kloesch_Strobl,Kummer_Widerin}, 
because our special boundary term (\ref{improvedsurfaceterm})
has enacted a canonical transformation. As we shall see, the $\Sigma${\sc cv}s 
are the proper ones to use 
when examining the correspondence between {\sc ecm}s and {\sc gdt}s, although
they are not ideal to take advantage of the underlying Poisson
structure associated with 
{\sc ecm}s.\cite{Schaller_Strobl,Kloesch_Strobl}


\subsection{Quasilocal Energy and Momentum}


Finding the canonical form of the action 
(\ref{correctPSMaction}) is not difficult. 
The volume-integral piece of 
(\ref{correctPSMaction}) is treated in the usual way; 
therefore, let us concentrate on how to handle the 
boundary terms. To begin with, convert the combined 
$\Sigma'$ and $\Sigma''$ integrals into a volume integral, 
and then perform an integration by parts on the 
$\rho \dot{X}'$ term in the volume integral in order to 
shift the radial derivative off of $\dot{X}$ and onto $\rho$, 
thereby creating another $\bar{\cal T}$ boundary integral. 
At the same time, the term $\dot{\rho} = 
\frac{1}{2}\partial_{t} \log(- e^{+}{}_{r}/e^{-}{}_{r})$ is
expanded. These simple manipulations yield
\begin{equation}
  \kappa \int^{\Sigma''}_{\Sigma'}\rho {\rm d}X
- \kappa \int_{\bar{\cal T}}\bar{\rho} {\rm d}X
=
  \kappa \int_{\cal M} {\rm d}^{2}x(
  {\textstyle \frac{1}{2}}X'
  \dot{e}^{+}{}_{r}/e^{+}{}_{r}
- {\textstyle \frac{1}{2}} X' 
  \dot{e}^{-}{}_{r}/e^{-}{}_{r}
- \rho' \dot{X})
- \kappa \int_{\bar{\cal T}}
  {\rm d}t\eta \dot{X}
{\,}.
\end{equation}
From this result the 
canonical form of the action follows:
\begin{equation}
  L = 
  \int_{\cal M}{\rm d}^{2}x 
  (P_{a} \dot{e}^{a}{}_{r} + P_{X}\dot{X}
- e^{a}{}_{t} G_{a} - \omega_{t} G) 
- \int_{\bar{\cal T}} {\rm d}t 
  (\kappa \eta \dot{X} + \bar{N} \bar{E})
{\,} ,
\label{canonicalaction1}
\end{equation}
with\footnote{For the particular case of {\sc ssgr}, Guven and
\'{O} Murchadha\cite{Guven_O_Murchadha}, have used what are 
essentially the gravitational constraints in this form  
(straightforwardly generalized to include matter) in an 
extensive examination of the spherically symmetric gravitational 
collapse of matter. This lightlike form of the constraints for {\sc ssgr}
had been earlier obtained by Malec and \'{O} 
Murchadha\cite{Malec_O_Murchadha}. One of the main achievements of
this work has been the derivation of certain bounds on the 
``optical scalars,'' essentially our $X^{\pm}$, during the collapse
scheme.}
\begin{eqnarray}
  G_{\pm}/\kappa & = & 
- \Lambda( n[X^{\mp}] \mp X^{\mp} \omega[n] \pm V n^{\mp})
\label{constraints0} \eqnum{\ref{constraints0}a} \\
  G/\kappa & = & 
- \Lambda( n[X] - X^{+} n^{-} + X^{-} n^{+})  
\eqnum{\ref{constraints0}b} \\
  \bar{E}/\kappa & = & X^{+} \bar{u}^{-} + X^{-} \bar{u}^{+}
{\,} . 
\eqnum{\ref{constraints0}c}
\addtocounter{equation}{1}
\end{eqnarray}
It is, of course, a trivial exercise to express $G_{\pm}$
in terms of the canonical variables 
$\{P_{\pm}; e^{\pm}{}_{r}; P_{X}, X\}$ but the resulting 
expressions are slightly more complicated and are not needed
here. Do note, however, that the $\Sigma${\sc cv}s are
tailored to  simplify $G = \Lambda (P_{-} n^{-} - P_{+} n^{+})$.
As is well-known\cite{Kloesch_Strobl,Kummer_Widerin}, $\{G_{a}, G\}$ 
comprise a set of first-class constraints, with associated 
Lagrange multipliers $\{e^{a}{}_{t}, \omega_{t}\}$. We would
like to point out that the structure of the boundary term
in the canonical action (\ref{canonicalaction1}) precisely 
mimics the corresponding boundary term in the canonical action
for full 4d general relativity, in particular
possessing a ``kinetic term'' with the parameter describing the 
boost between $u$ and $\bar{u}$ (slice and boundary 
frames) playing the role of a 
``momentum''\cite{GHayward_Wong,Lau2}. 

We interpret the expression for $\bar{E}$ as the 
{\em quasilocal energy}\footnote{The terminology comes from 
{\sc ssgr}, where this energy is associated with a round sphere, 
which is identified with a point in our $1+1$ spacetime 
region ${\cal M}$. $\bar{E}$ is certainly {\em not} an energy
{\em density} like, for example, the component $T^{t}_{t}$ of the
matter energy-momentum tensor.} 
({\sc qle}) associated with the $\bar{u}^{\mu}$ 
observers at the $\bar{\cal T}$ boundary. It is a convenient 
and standard practice to associate a separate energy with 
each boundary point\cite{Witten}, with the {\em expression} for the 
energy at an inner boundary point being minus the 
{\em expression} for the energy at an outer boundary point.
(We emphasize that, although the inner and outer boundary-point 
{\em expressions} for the energy differ by a sign, minus
the {\em value} of the inner boundary-point energy is not equal
to the {\em value} of outer boundary-point energy.) For instance, 
$\bar{E}|_{B_{o}}$ is the energy associated with the outer 
boundary point. Put precisely, $\bar{E} |_{B_{o}}$ is the energy, 
associated with $B_{o}$, of the gravitational fields (and possibly 
minimally coupled matter fields, if they are present)
which ``live'' on a spacelike slice $\bar{\Sigma}$, 
where $\bar{\Sigma}$ has $\bar{u}^{\mu}$ as its timelike 
normal at the point 
$B_{o} = \partial \bar{\Sigma}$.\footnote{There is
an equivalence class of such $\bar{\Sigma}$'s (of course, technically
$\partial \bar{\Sigma}$ is the union $B_{i} \bigcup B_{o}$).}
Inspection of the alternative 
variation (\ref{alternatevariation}) 
of the action verifies the following important statement: 
When evaluated on solutions to the {\sc eom}s, the energy
$\bar{E}|_{B_{o}}$, for example, is minus the time rate of 
change of the classical action (or Hamilton-Jacobi principal 
function) with respect to a unit stretch in 
$\bar{N}|_{\bar{\cal T}_{o}}$, where $\bar{N}|_{\bar{\cal T}_{o}}$ 
controls the lapse of proper time between neighboring points on 
$\bar{\cal T}_{o}$. For definiteness
in what follows, we shall confine ourselves to the outer 
boundary-point expression for the energy.

The {\sc qle} of the physical fields defined on the slice $\Sigma$ and 
associated with the $u^{\mu}$ observers at the boundary 
$B_{o} = \partial\Sigma$ is
\begin{equation}
E |_{B_{o}} = \kappa (X^{+} u^{-} + X^{-} u^{+}) {\,}.
\label{naiveQLE}
\end{equation}
(Henceforth, we shall suppress all $|_{B_{o}}$'s, with the 
understanding the we only deal with the outer boundary-point 
expression.) Now, any suitable expression for the $\Sigma$ energy
should be expressible solely in terms of the 
canonical variables of $\Sigma$, and this energy satisfies this
requirement. Indeed, using the first of the important 
geometric identities\footnote{Respectively,
(\ref{geometricidentity}a) and (\ref{geometricidentity}b) 
can be written as the manifestly zweibein-gauge 
invariant expressions $- X^{a} u^{b} \eta_{ab} = X^{a} n^{b} 
\epsilon_{ab}$ and $- X^{a} n^{b} \eta_{ab} = X^{a} u^{b}\epsilon_{ab}$.} 
\begin{eqnarray}
X^{+} u^{-} +
X^{-} u^{+} & = & X^{-} n^{+} - X^{+} n^{-}
{\,},
\label{geometricidentity}
\eqnum{\ref{geometricidentity}a} \\
X^{+} n^{-} + X^{-} n^{+} & = & X^{-} u^{+} - X^{+} u^{-}
\eqnum{\ref{geometricidentity}b}
\addtocounter{equation}{1}
\end{eqnarray}
we write
\begin{equation}
E = \kappa (X^{-} n^{+} - X^{+} n^{-})
{\, }. 
\label{QLE}
\end{equation} 
Again, we could easily write the $X^{\pm}, n^{\pm}$ 
here in terms of the chosen canonical variables, but choose not to.
Despite its zweibein-gauge invariance 
[cf.~the footnote just before (\ref{geometricidentity})],
the {\sc qle} is clearly foliation-dependent, because the 
foliation choice determines $n^{\mu}$. 
We shall need a {\sc qle} expression for each point of $\Sigma$. Fixing
the sign of the energy by interpreting each $\Sigma$ point as
an outer-boundary point for a portion of $\Sigma$, we can use
the expression (\ref{QLE}) as given. Notice that, modulo the 
Lorentz constraint $G$, the {\sc qle} has the form
\begin{eqnarray}
  E & = &  - \kappa n[X] \nonumber \\
    & \uparrow & {\rm equality\,\,modulo}\,\,G\,\,
                 [{\rm or\,\,equivalently\,\,modulo\,\,the}\,\,
                 {\sc eom}\,\, (\ref{eom}{\rm c})].
\label{onshellQLE}
\end{eqnarray}
Furthermore, this result immediately implies that the correct 
expression for {\sc qle} in the conformally rescaled spacetime,
that is the physical energy, is
\begin{equation}
\tilde{E} = \exp (\varphi) E 
{\,} ;
\label{tildeQLE}
\end{equation}
and, therefore, the expression 
for {\sc qle} in {\sc gdt}s is simply $\tilde{E} =
- \kappa \tilde{n}[X]$. Again, as a vector field $\tilde{n}^{\mu} = 
\tilde{\Lambda}^{-1} \delta^{\mu}_{r}$, and in fact 
$\tilde{n}^{\mu} = \exp (\varphi) n^{\mu}$.
Finally, we would like to point out that the on-shell expression
(\ref{onshellQLE}) for the energy is in fact {\em negative} in an
(exterior) type I region, as for the classical solutions we 
consider $X' > 0$ in such a region (and as
a result the $\tilde{E}$ in (\ref{tildeQLE}) is also negative in
a type I region). 
One obtains a positive energy expression only after referencing the 
energy as given against the groundstate energy of the theory. 
We address this issue in detail in $\S$ VI.C.

We also define
\begin{equation}
J = - \kappa (X^{+}n^{-} + X^{-}  n^{+})
\label{J} 
\end{equation}
and consider $J$, which is nothing but $-P_{\Lambda}$, to
be a {\em quasilocal local momentum}. Trivially, $J$ depends only
on $\Sigma$ canonical variables. Notice that on-shell
$J |_{\Sigma''}$, for instance, is minus the rate of change
of the classical action with respect to a unit-stretch in 
$\Lambda |_{\Sigma''}$, where $\Lambda |_{\Sigma''}$ controls
the lapse of proper radial distance between neighboring points
of $\Sigma''$. Again, inspection of (\ref{alternatevariation})
verifies this statement. Next, with the geometric identity
(\ref{geometricidentity}b)
and the {\sc eom} (\ref{eom}c), one sees that
the on-shell expression for $J$ is
\begin{eqnarray}
J & = & \kappa u[X] \nonumber \\
  & \uparrow & {\rm equality\,\,modulo\,\,the}
               \,\,{\sc eom}\,\,(\ref{eom}{\rm c}).
\end{eqnarray}
The conformally rescaled expression is $\tilde{J} = 
\exp(\varphi) J$, and, hence, the expression for quasilocal
momentum in {\sc gdt} is simply $\tilde{J} = \kappa \tilde{u}[X]$.
We can also define a quasilocal momentum $\bar{J}$ for the 
observers in the $\bar{\cal T}$ boundary,
\begin{equation}
\bar{J} = - \kappa (X^{+}\bar{n}^{-} + X^{-} \bar{n}^{+}) 
{\,} .
\label{Jbar}
\end{equation} 
Eqs.~(\ref{boundaryboost}) imply the following boost 
relations between the ``barred'' and ``unbarred''
quasilocal energy and momentum:
\begin{eqnarray}
\bar{E} & = & \gamma E - v \gamma J  
\label{boostedQLE} \eqnum{\ref{boostedQLE}a} \\
\bar{J} & = & \gamma J - v \gamma E 
\eqnum{\ref{boostedQLE}b}
{\,} .
\addtocounter{equation}{1}
\end{eqnarray}
Up to now, our results concerning quasilocal energy-momentum are 
unaffected by the inclusion of 2d minimally coupled matter. 

We now turn to the special case when the spacetimes under 
consideration possess an isometry (as is always true in 
vacuo).
Just like before, we define quasilocal energy and momentum 
expressions for the preferred foliation $\hat{\Sigma}$ 
associated with the Killing direction, 
[cf.~Eq.~(\ref{killingboost})]
\begin{eqnarray}
  \hat{E} &  = & 
  {\textstyle \frac{1}{2}}\kappa (\sigplus + \sigminus) 
  (X^{-} \hat{n}^{+} - X^{+} \hat{n}^{-}) 
+ {\textstyle \frac{1}{2}}\kappa
  (\sigplus - \sigminus)
  (X^{-} \hat{u}^{+} - X^{+} \hat{u}^{-})
\label{KillingQLEM} \eqnum{\ref{KillingQLEM}a} \\
- \hat{J} & = & 
  {\textstyle \frac{1}{2}}\kappa (\sigplus + \sigminus)
  (X^{+} \hat{n}^{-} + X^{-} \hat{n}^{+}) 
+ {\textstyle \frac{1}{2}}\kappa (\sigplus - \sigminus)
(X^{+} \hat{u}^{-} + X^{-} \hat{u}^{+})
{\,} .
\eqnum{\ref{KillingQLEM}b}
\addtocounter{equation}{1}
\end{eqnarray}
Notice the appearance of $\sigma_{\pm}$ in these expressions, 
which is necessary to cope with the fact that for the
pathological $\hat{\Sigma}$ foliation the roles of $\sigplus\hat{u}$ and
$\sigminus\hat{n}$ switch at the horizon. With the exception of
the horizon where $\sigK$ is not defined, the relations 
(\ref{killingboost}) along with the identities 
(\ref{geometricidentity}a,b) 
determine
\begin{eqnarray}
\hat{E} & = & \hat{\gamma} E - \hat{v}\hat{\gamma} J  
\label{boostedQLE2} \eqnum{\ref{boostedQLE2}a} \\
\hat{J} & = & \hat{\gamma} J - \hat{v} \hat{\gamma} E 
\eqnum{\ref{boostedQLE2}b}
\addtocounter{equation}{1}
\end{eqnarray}
as the associated boost relations between ``hatted'' and ``unhatted''
quasilocal energy and momentum. Now, in fact, the ``hatted'' versions
of the identities (\ref{geometricidentity}a,b) in tandem 
with the {\sc eom} (\ref{eom}c) tell 
us that in a static region $\hat{J} 
= \kappa \sigma_{+}\hat{u} [X] = 0$, while in a dynamical region 
$\hat{E} = - \kappa \sigma_{+}\hat{u}[X] = 0$. The last equalities
follow since $\hat{u}^{\mu}$ points along the Killing direction. 
Therefore, by the boost relations inverse to (\ref{boostedQLE2}), we 
know that on-shell the {\em rapidity} boost parameter
\begin{eqnarray}
  \hat{\eta} = {\textstyle \frac{1}{2}} 
  \log \left|\frac{E + J}{E - J}\right|
& = & {\textstyle \frac{1}{2}} \log \left| 
\frac{1 + \hat{v}}{1 - \hat{v}}\right| \nonumber \\
& \uparrow & {\rm equality\,\,modulo\,\,the}
               \,\,{\sc eom}\,\,(\ref{eom}{\rm c}).
\end{eqnarray}
describes the local boost between $u^{\mu}$ and 
$\sigplus \hat{u}^{\mu}$ (or
$\sigminus \hat{n}^{\mu}$ in dynamical region), that is between the 
$\Sigma$ and $\hat{\Sigma}$ frames. We have, therefore,
an expression in terms of the $\Sigma$ canonical variables
for the boost parameter introduced and discussed in $\S$ II
(in the non-vacuum case it certainly may not be possible to 
obtain such an expression).
A straightforward calculation
shows that
\begin{equation}
  \hat{\eta} = - {\textstyle \frac{1}{2}} \log (- n^{+}/n^{-}) 
  + {\textstyle \frac{1}{2}} \log |X^{+}/X^{-}|
{\,} ;
\end{equation}
or, in other words, that $\xi = \rho + \hat{\eta}$. Here for 
later purposes we define a new boost parameter
\begin{equation}
\xi := {\textstyle \frac{1}{2}} \log| X^{+}/X^{-}|
{\,},
\label{firstrapid}
\end{equation}
also to be called a {\em rapidity}.
Clearly, the parameter $\xi$ is $\mp \infty$ if $X^{\pm} = 0$ 
vanishes while $X^{\mp}$ does not. Indeed, as is evident from
the identity
\begin{equation}
X^{+}/X^{-} = \sigK \exp(\alpha X)|C - W|^{-1} 
(X^{+})^{2}
{\,} ,
\label{ratioXplusXminus}
\end{equation}
this happens at the horizon. As one might expect, the 
limiting value of $\xi$, like $\hat{\eta}$, at a bifurcation
point in a Kruskal-type diagram is direction-dependent.

\section{Canonical Transformations and Rest-Frame Energy}
In this section we have several goals in mind. 
First, we provide
a rigorous derivation of {\sc gdt} from the {\sc ecm} formalism. 
Second, we obtain the preferred canonical variables 
(for both {\sc ecm}s and {\sc gdt}s) which are associated with the 
underlying Poisson structure (implicit in the {\sc ecm}s we 
consider). 
The preferred canonical variables we obtain are essentially 
the ``Casimir-Darboux coordinates'' used by Kl\"{o}sch and
Strobl\cite{Kloesch_Strobl}. 
For reasons which will be evident, we call our 
preferred canonical variables ``rapidity 
canonical variables'' ({\sc rcv}s).
Third, having found the {\sc rcv}s, we pass to the  
reduced phase space of the theory by elimination of the 
constraints, in order to obtain the ``true degrees of freedom'' 
for both (vacuum) {\sc ecm}s and {\sc gdt}s simultaneously. Fourth, we 
use the {\sc rcv}s to examine the concept of 
``rest-frame'' energy for {\sc gdt}s with a Minkowski-spacetime 
ground state. Throughout this section we work in vacuo, 
although the derivation of {\sc gdt} from the {\sc ecm} formalism is 
unaffected by the inclusion of matter coupled in a not too ``exotic''
manner.

\subsection{Derivation of GDTs from ECMs}

As is evident from the alternative variation 
(\ref{alternatevariation})
of the action, the following is a trivial 
canonical transformation:
\begin{equation}
  \{P_{\pm}, e^{\pm}{}_{r}; P_{X}, X\} \to 
  \{P_{\Lambda}, \Lambda;  P_{X}, X; P_{\rho},\rho\}
{\, }.
\label{timegaugect}
\end{equation}
By ``trivial,'' we mean that the transformation requires 
{\em no} generating
functional.
Along with this canonical transformation, one should also 
recombine the constraints. 
(Although, depending on one's ultimate
goal, this may be work done now 
which has to be undone later!) 
The proper recombination is easily 
achieved by inserting the identities  
$e^{\pm}{}_{t} = \sqrt{\frac{1}{2}} \sigL e^{\pm\rho}(
N \pm \Lambda N^{r})$ into the canonical action 
(\ref{canonicalaction1}), with the result
\begin{equation}
  L = \int_{\cal M} {\rm d}^{2}x 
  (P_{\Lambda} \dot{\Lambda} 
+ P_{X} \dot{X} 
+ P_{\rho} \dot{\rho} 
- N {\cal H} 
- N^{r} {\cal H}_{r} 
- \omega_{t} G) 
- \int_{\bar{\cal T}}{\rm d}t
  (\kappa \eta \dot{X} 
+ \bar{N} \bar{E})
{\,} ,
\end{equation}
where $E$ and $J$ (found in $\bar{E} 
= \gamma E - v\gamma J$), $G$, 
and the new constraints 
\begin{eqnarray}
  {\cal H} & = & \sqrt{\textstyle \frac{1}{2}} \sigL
  (e^{\rho} G_{+} + e^{-\rho} G_{-})
\label{standardconstraints} \eqnum{\ref{standardconstraints}a}\\
  {\cal H}_{r} & = & \sqrt{\textstyle \frac{1}{2}} \sigL \Lambda
  (e^{\rho} G_{+} - e^{-\rho} G_{-})
\eqnum{\ref{standardconstraints}b} 
\addtocounter{equation}{1}
\end{eqnarray}
must be expressed in terms of 
$\{P_{\Lambda}, \Lambda;  P_{X}, X; P_{\rho},\rho\}$. In fact,
as is evident from (\ref{alternatevariation}), one already has
$P_{\rho} = -G$. A straightforward calculation yields 
the remaining 
set
\begin{eqnarray}
{\cal H} & = &  - \kappa^{-1} P_{\Lambda} P_{X} 
+ \left[\Lambda^{-1}(\kappa X' - P_{\rho})\right]'
\nonumber \\
& & 
+ {\textstyle \frac{1}{2}} \alpha \kappa^{-1}
  \left[\Lambda^{-1}(\kappa X' - P_{\rho})^{2} 
- \Lambda(P_{\Lambda})^{2}\right] 
+ \kappa\lambda^{2} \Lambda V_{0} 
\label{standardconstraints2} 
\eqnum{\ref{standardconstraints2}a} \\
{\cal H}_{r} & = & 
- \Lambda (P_{\Lambda})' + P_{X} X' 
- \kappa^{-1} P_{X} P_{\rho} 
\eqnum{\ref{standardconstraints2}b} \\
\bar{E} & = & 
- \gamma\Lambda^{-1}(\kappa X' - P_{\rho}) 
+ v\gamma P_{\Lambda}
\eqnum{\ref{standardconstraints2}c}
\addtocounter{equation}{1}
\end{eqnarray}
Since the pair $\{P_{\rho}, \rho\}$ is ``pure gauge,''  
it may be eliminated completely (set to zero in all equations) 
without affecting the {\em solutions} of the constraints and 
equations of motion for the remaining canonical 
variables. The elimination of this kinematic phase-space pair 
corresponds, at the covariant level, to solving the 
{\sc eom} (\ref{eom}c) and substituting the solution back 
into the {\sc ecm} action. We wish to stress that our ability to eliminate 
this gauge pair is intimately related to the particular boundary term
(\ref{improvedsurfaceterm}) that we have added to the {\sc ecm} action.
(Of course, a similar elimination of this gauge pair would have been 
possible had we added {\em no} boundary integral at all to the {\sc ecm} 
action; however, were we to follow that route, we would encounter
an obstacle when trying to perform the ``conformal'' canonical 
transformation below.) The result of this elimination is a 
perfectly good canonical form 
of the theory based on $\{P_{\Lambda},\Lambda; P_{X}, X\}$. 

Now, recall that, at the covariant level, a step in the 
passage from {\sc ecm}s to {\sc gdt}s involves the conformal rescaling 
discussed in $\S$ I.C\cite{KKL1}. At the canonical level this conformal 
rescaling is mirrored as a ``trivial'' canonical 
transformation subsequent\footnote{Actually, the ``conformal'' 
canonical transformation can be performed prior to the transformation 
(\ref{timegaugect}) (and hence prior to elimination of
the gauge pair $\{P_{\rho},\rho\}$).} 
to the first one (\ref{timegaugect}),
\begin{equation}
  \{P_{\Lambda},\Lambda; P_{X}, X\} \to 
  \{{\cal P}_{\tilde{\Lambda}},\tilde{\Lambda}; {\cal P}_{X},X\}
{\,} ,
\label{tildecanonicaltransformation}
\end{equation}
where the new canonical variables are
\begin{eqnarray}
  {\cal P}_{\tilde{\Lambda}} & = & 
\exp (\varphi) P_{\Lambda}
\label{metricmomenta} 
\eqnum{\ref{metricmomenta}a} \\
  \tilde{\Lambda} & = & \exp (-\varphi) \Lambda
\eqnum{\ref{metricmomenta}c} \\
  {\cal P}_{X} & = & 
  P_{X} + ({\rm d}\varphi/{\rm d}X)\Lambda P_{\Lambda} 
{\,} .
\eqnum{\ref{metricmomenta}c}
\addtocounter{equation}{1}
\end{eqnarray}
Of course, the choice of conformal factor is model-dependent.
Under this canonical transformation, we obtain the canonical
action corresponding to a {\sc gdt}, namely,
\begin{equation}
  \tilde{L} = \int_{\cal M} {\rm d}^{2}x 
  ({\cal P}_{\tilde{\Lambda}} \dot{\tilde{\Lambda}}
+ {\cal P}_{X} \dot{X} 
- \tilde{N} \tilde{\cal H} 
- N^{r} {\cal H}_{r}) 
- \int_{\bar{\cal T}} {\rm d}t 
  (\kappa \eta \dot{X} + \tilde{\bar{N}} \tilde{\bar{E}})
\label{ADMaction}\end{equation}
with the following expressions for the constraints and quasilocal
energy:
\begin{eqnarray}
\tilde{\cal H} & = & 
- \kappa^{-1}{\cal P}_{\tilde{\Lambda}} {\cal P}_{X} 
+ \kappa^{-1}
  \left({\rm d} \varphi^{\scriptscriptstyle NEW}/{\rm d}X\right) 
  \tilde{\Lambda} 
  ({\cal P}_{\tilde{\Lambda}})^{2}
\nonumber \\
& &
+ \kappa\exp \left(\varphi^{\scriptscriptstyle NEW}\right) 
  \left[\exp \left(-\varphi^{\scriptscriptstyle NEW}\right)
  \tilde{\Lambda}{}^{-1}X'\right]'
+ \kappa \lambda^{2} \exp (2\varphi) \tilde{\Lambda} V_{0}
\label{rescaledconstraints} \eqnum{\ref{rescaledconstraints}a}\\
  {\cal H}_{r} & = & - \tilde{\Lambda} 
  ({\cal P}_{\tilde{\Lambda}})' + {\cal P}_{X} X'
\eqnum{\ref{rescaledconstraints}b} \\
\tilde{\bar{E}} & = & - \gamma\tilde{\Lambda}^{-1}\kappa X' 
              + v\gamma {\cal P}_{\tilde{\Lambda}} 
{\,} .
\eqnum{\ref{rescaledconstraints}c}
\addtocounter{equation}{1}
\end{eqnarray}
Here we have defined the ``new'' conformal factor\footnote{Here, 
for once, we have dropped our previous restriction to 
$\alpha =$ const.~in (\ref{PSMpotential}), i.~e.~we 
allow for the general case $\alpha = \alpha (X)$.}
\begin{equation}
  \varphi^{\scriptscriptstyle NEW} := 
  \varphi - {\textstyle \frac{1}{2}}
  \int^{X}_{X_{0}}{\rm d}y 
  {\,}\alpha(y)
\label{new_factor}
\end{equation}
by subtracting a field integral of $\alpha$ from the 
``old'' conformal factor. 
Notice that the lapse $\tilde{N}$, boundary lapse 
$\tilde{\bar{N}}$, Hamiltonian constraint $\tilde{\cal H}$, 
and quasilocal energy $\tilde{\bar{E}} = \gamma \tilde{E} 
- v\gamma \tilde{J}$ are 
rescaled quantities, while the shift $N^{r}$ and momentum 
constraint ${\cal H}_{r}$ are not (although, of course,
${\cal H}_{r}$ must still be re-expressed in terms of the 
new ``tilded'' canonical variables). 

The ability to perform the canonical transformation
(\ref{tildecanonicaltransformation}) renders the 
function $\alpha$ {\em locally} irrelevant (although vastly
different global properties can stem from changes in 
$\alpha$ and from the conformal rescaling\cite{KKL1}!).
Indeed, because $V_{0} = V_{0}(X)$ is essentially arbitrary
in our formalism, we may rewrite the potential term 
$V_{0}$ in $\tilde{\cal H}$ as 
\begin{equation}
  V_{0} = \exp\left[- \int^{X}_{X_{0}}
  {\rm d}y{\,} \alpha(y)\right] 
V^{\scriptscriptstyle NEW}_{0}
{\,} .
\end{equation} 
Then, we can substitute $\exp(2\varphi^{\scriptscriptstyle NEW})
V^{\scriptscriptstyle NEW}_{0}$ for $\exp(2\varphi) V_{0}$  in
the right-most term in (\ref{rescaledconstraints}a).
This shows that $\tilde{\cal H}$ may be obtained from
$\cal H$ above with $\alpha = 0$ by using the new conformal 
factor (\ref{new_factor}) and choosing a slightly different 
potential $V_{0}$. This is our argument as to why our choice of 
$\alpha = 0$ in $\S$ IV results in no loss of generality at the
level of {\sc gdt} as long as ${\cal M}$ is not ``cut'' but the
conformal transformation.\footnote{A positive (or negative) conformal
factor typically produces new geodesically complete edges inside
the global Penrose diagram of the corresponding {\sc ecm} \cite{KKL1}.}
Finally, if desired, we may make a yet another trivial canonical 
transformation in order to trade the pair $\{{\cal P}_{X},X\}$ 
for the new pair $\{{\cal P}_{\tilde{X}}, \tilde{X}\}$, with 
$\tilde{X}$ the dilaton field of $\S$ IV and ${\cal P}_{\tilde{X}} 
= 2\lambda^{-2}|W^{\scriptscriptstyle NEW}|{\cal P}_{X}$ (here 
$W^{\scriptscriptstyle NEW}$ has the same definition as $W$ but 
with $V_{0}^{\scriptscriptstyle NEW}$). The result of our labors is
the canonical form of the action (\ref{hatdilatonaction2})
[provided, of course, that in 
(\ref{hatdilatonaction2}) the zweibein as been suitably gauge-fixed 
at the boundary so that (\ref{hatdilatonaction2}) is truly a metric 
action]. 

For the special cases of 
{\sc ssgr} and {\footnotesize 2d}{\sc dg} in the presence of a Killing
field, the {\em base} piece of the 
canonical {\sc adm} action (\ref{ADMaction}) above is the 
starting point for many 
authors\cite{Kuchar,Louko_Whiting,Lau,Varadarajan,Bose_Louko_Parker_Peleg},
who then essentially work in a direction opposite to our course in
this subsection with the ultimate goal of obtaining the ``true 
degrees of freedom,'' via a canonical transformation of the form 
$\{{\cal P}_{\tilde{\Lambda}},\tilde{\Lambda}; {\cal P}_{X}, X\} \to 
\{{\sf P}_{C}, C; {\sf P}_{X}, X\}$. Again $C$ is our absolutely
conserved quantity which corresponds in metric-variable {\sc ssgr} 
to Kucha\v{r}'s canonical expression for the {\sc sbh} mass 
parameter\cite{Kuchar}. Such a transformation and mass canonical variable
have also been considered for {\sc gdt} in 
Refs.~\cite{Barvinsky_Kunstatter}. 
In the next subsection we shall obtain the pair 
$\{{\sf P}_{C}, C; {\sf P}_{X}, X\}$ in a 
more direct fashion from the {\sc ecm} canonical variables.

\subsection{Rapidity Canonical Variables}
 
We now perform two successive canonical transformations on 
the $\Sigma${\sc cv}s $\{P_{a}, e^{a}{}\,_{r}; P_{X}, X\}$ 
in order to obtain the {\sc rcv}s which clearly exhibit the
true degrees of freedom for the models we study. Also 
examined in detail is the generating functional for the 
composite transformation. 
The first canonical transformation 
$\{P_{a}, e^{a}{}_{r}; P_{X},X\}
\to \{{\cal P}_{a}, e^{a}{}_{r};{\cal P}_{X}, X\}$ is to the 
standard (Poisson-sigma model) canonical 
variables of an 
{\sc ecm}:\cite{Schaller_Strobl,Kloesch_Strobl,Kummer_Widerin}
${\cal P}_{\pm} = \kappa X^{\mp}$ and ${\cal P}_{X} = 
- \kappa\Lambda \omega[n]$. That this transformation is 
indeed canonical is verified by the identity
\begin{equation}
  P_{a} \delta e^{a}{}_{r} + P_{X} \delta X - {\cal P}_{a} 
  \delta e^{a}{}_{r} - {\cal P}_{X} \delta X 
= \kappa \delta (\rho X') - \kappa (\rho \delta X)'
{\,} ,
\label{identityone}
\end{equation}
which upon integration over a generic slice $\Sigma$ shows that,
{\em for our boundary conditions} (that is, fixation of $X$), 
the difference between the old and new Liouville forms is an exact
form. Hence, the transformation is canonical. We now perform 
a second canonical transformation to {\em rapidity canonical variables} 
[so-named since the rapidity variable (\ref{firstrapid}) appears as a 
phase-space coordinate]. Define the new canonical variables as\footnote{In 
(\ref{cdcoords}) and some of the equations to follow, the 
purist will wish to express $X^{\pm}$, $\Lambda n^{\pm}$, 
$\Lambda \omega[n]$, and $\Lambda n[X]$ in terms of either
$\{{\cal P}_{a}, e^{a}{}_{r}, {\cal P}_{X}, X \}$ or, as the 
case may be, the original canonical variables 
$\{P_{a}, e^{a}{}_{r}; P_{X},X\}$; however, as mentioned, 
in order to keep the expressions as simple as possible, we 
shall not do so.} [cf.~Eq.~(\ref{W_definition})]
\begin{eqnarray}
  C & = & \exp (\alpha X) X^{+} X^{-} + W
\label{cdcoords} \eqnum{\ref{cdcoords}a} \\
  {\sf P}_{C} & = & 
- {\textstyle \frac{1}{2}}  \kappa\Lambda\exp (-\alpha X)
  \left[(n^{+}/X^{+}) 
+ (n^{-}/X^{-})\right]
\eqnum{\ref{cdcoords}b} \\
  \xi & = & {\textstyle \frac{1}{2}} \log 
  \left|X^{+}/X^{-}\right|
\eqnum{\ref{cdcoords}c} \\
  {\sf P}_{\xi} & = & \kappa \Lambda\left( n[X] 
  - n^{-} X^{+} + n^{+}X^{-}\right)
\eqnum{\ref{cdcoords}d} \\
X & = & X 
\eqnum{\ref{cdcoords}e} \\
  {\sf P}_{X} & = & - \kappa \Lambda\left\{\omega[n] 
- {\textstyle \frac{1}{2}}
  V\left[(n^{+}/X^{+}) 
+ (n^{-}/X^{-})\right] 
+ {\textstyle \frac{1}{2}}
  n\left[\log\left|X^{+}/X^{-}\right|\right]\right\}
{\,} .
\eqnum{\ref{cdcoords}f}
\addtocounter{equation}{1}
\end{eqnarray}
Clearly, this transformation breaks down on a horizon. 
However, for the time being, we 
shall ignore this fact. 
Note that ${\sf P}_{\xi} = - G$; and, therefore, 
one expects the pair $\{ {\sf P}_{\xi},\xi\}$ to be 
``pure gauge.'' This expectation turns out to be correct.
A comparison with the approach of Kl\"{o}sch 
and Strobl\cite{Kloesch_Strobl} is in
order. They work with coordinates in which
only $\log |X^{+}|$ (or $\log |X^{-}|$, but not the 
logarithm of the ratio) and its conjugate appear as one of 
the phase-space pairs. 
This has advantages and disadvantages.
The former being that a given set of  Kl\"{o}sch and Strobl's 
canonical variables are good on part of the horizon, in marked 
contrast to our variables which are bad on the whole horizon. 
However, our {\sc rcv}s afford easy passage to metric {\sc gdt}.

To establish that the composite transformation 
$\{P_{\pm}, e^{\pm}{}_{r} ; P_{X}, X\} \to 
\{{\sf P}_{C} , C; {\sf P}_{X} , X; {\sf P}_{\xi} , \xi\}$
from $\Sigma${\sc cv}s to {\sc rcv}s is canonical, first 
verify the following identity:
\begin{equation}
{\cal P}_{a} \delta e^{a}{}_{r} + {\cal P}_{X}\delta X
- {\sf P}_{C} \delta C - {\sf P}_{\xi} \delta \xi 
- {\sf P}_{X} \delta X = - \kappa \delta (X_{a} e^{a}{}_{r} 
+ \xi X') + \kappa (\xi \delta X)'
{\,} ,
\label{identitytwo}
\end{equation}
which when combined with (\ref{identityone}) gives a third
identity
\begin{equation}
P_{a} \delta e^{a}{}_{r} + P_{X}\delta X
- {\sf P}_{C} \delta C - {\sf P}_{\xi} \delta \xi 
- {\sf P}_{X} \delta X = - \kappa \delta [X_{a} e^{a}{}_{r} 
+ (\xi - \rho) X'] + \kappa[(\xi -\rho)\delta X]'
{\,} .
\label{identitythree}
\end{equation}
Integration of (\ref{identitythree}) 
over a generic slice $\Sigma$
yields
\begin{eqnarray}
\lefteqn{\int_{\Sigma}{\rm d}r (P_{a} 
  \delta e^{a}{}_{r} + P_{X}\delta X)
- \int_{\Sigma} {\rm d}r ({\sf P}_{C} 
  \delta C + {\sf P}_{\xi} 
  \delta \xi + {\sf P}_{X} \delta X) =} 
& & \nonumber \\
& & 
-  \kappa \delta \int_{\Sigma} {\rm d}r
  [{X}_{a} e^{a}{}_{r} 
+ (\xi - \rho) X'] + \Bigl. 
  \kappa[(\xi -\rho)\delta X]\Bigr|^{B_{o}}_{B_{i}}
{\,} ;
\label{identityfour}
\end{eqnarray}
therefore, for our boundary conditions (that is, with 
fixation of $X$ at the boundary $\partial \Sigma = 
B_{i} \bigcup B_{o}$) the 
difference between the old and new Liouville forms is an 
exact form. Hence, the composite transformation 
is indeed canonical, although to be performed it requires a 
generating functional, namely,
\begin{equation}
  \Xi |_{\Sigma} := \kappa \int_{\Sigma} 
  {\rm d}r[X_{a} e^{a}{}_{r} +
  (\xi - \rho) X']
{\,} .
\label{generatingfunctionalfirst}
\end{equation}
Using the results of $\S$ V, we easily see that the generating
functional depends on the quasilocal energy and momentum,
\begin{equation}
\Xi |_{\Sigma}  = \int_{\Sigma} {\rm d}r \Lambda[ J - \hat{\eta} E 
- \hat{\eta} G/\Lambda]
{\,} .
\label{generatingfunctional}
\end{equation}
We could, of course, work with
a slightly modified generating functional 
\begin{equation}
\Upsilon |_{\Sigma} := \int_{\Sigma} {\rm d}r \Lambda[J - \hat{\eta} E]
\end{equation}
in place of $\Xi |_{\Sigma}$. 
This is permissible, as using $\Upsilon$ amounts to the 
addition of an {\sc eom} 
boundary term to the action, i.~e.~a term  which vanishes 
on-shell. In this case, if one is willing to redefine the 
Lagrange parameter associated with the Lorentz constraint, then 
$\Upsilon |_{\Sigma}$ gives rise to slightly different new canonical 
variables, namely, 
$\{{\sf P}_{C}, C; {\sf P}_{X} , X; {\sf P}_{\rho} 
= {\sf P}_{\xi} , \rho\}$. But, so far we see no particular
advantage to be gained by working with $\Upsilon |_{\Sigma}$ 
in place of $\Xi |_{\Sigma}$.

By replacing the arbitrary variation $\delta$ in (\ref{identityfour})
with a time derivative and then integrating the whole equation over 
time, we find
\begin{equation}
\int_{\cal M}{\rm d}^{2}x (P_{a} \dot{e}^{a}{}_{r} + P_{X} \dot{X})
+ \Xi|^{\Sigma''}_{\Sigma'} = \int_{\cal M}{\rm d}^{2}x
  ({\sf P}_{C} \dot{C} + {\sf P}_{X} \dot {X} + {\sf P}_{\xi} \dot{\xi})
+ \int_{\bar{\cal T}} {\rm d}t \kappa \hat{\eta} \dot{X} 
{\, } .
\label{new_p's_and_q's}
\end{equation}
Hence, defining a new action $L_{\dagger} = L + \Xi|^{\Sigma''}_{\Sigma'}$, 
with (\ref{new_p's_and_q's}) one arrives at\footnote{In 
Ref.~\cite{Lau} a step essentially identical to this one is handled 
poorly in obtaining equation (4.24) of that reference. 
Nevertheless, a heuristic argument in the text just after this 
equation compensates for this poor handling; hence, the analysis in 
that reference is correct.} 
\begin{equation}
L_{\dagger} =  \int_{\cal M}{\rm d}^{2}x({\sf P}_{C} \dot{C} 
+ {\sf P}_{X} \dot {X} + {\sf P}_{\xi} \dot{\xi} - e^{a}{}_{t} G_{a} 
- \omega_{t} G)
- \int_{\bar{\cal T}} {\rm d}t[ \kappa (\eta - \hat{\eta}) \dot{X}
+ \bar{N} \bar{E}]  
\label{daggeraction}
\end{equation}
for the canonical form of $L_{\dagger}$ [cf.~with the action
given in Eq.~(\ref{canonicalaction1})].
Let us express the constraints and quasilocal energy in 
(\ref{daggeraction}) in terms of the new canonical variables.
It is trivial to obtain the new expression for the 
Lorentz constraint; however, one must exercise due caution with 
the signs of $\sigma_{\pm}$,
$\sigK$ when obtaining the new expressions for 
$G_{\pm}$. Checking the identities
\begin{eqnarray}
- \kappa C' & = & 
  \exp (\alpha X )(X^{a} G_{a} + V G) 
\label{WKidentities} \eqnum{\ref{WKidentities}a} \\
  {\sf P}_{X} & = & {\textstyle \frac{1}{2}}
  \left[(G_{-}/X^{+}) - (G_{+}/X^{-})\right]
{\,} 
\eqnum{\ref{WKidentities}b}
\addtocounter{equation}{1}
\end{eqnarray}
is straightforward. Indeed, the first is by now a well-known 
result\cite{Martinez_Kunstatter,Fischler,Tada_Uehara,Kloesch_Strobl,Kummer_Widerin}.
From (\ref{WKidentities}a) and (\ref{cdcoords}c) we determine that
\begin{equation}
  G_{\pm} + \sigK e^{\mp 2\xi} G_{\mp} = (X^{\pm})^{-1}
  \left[V {\sf P}_{\xi} - \kappa \exp(-\alpha X)C'\right]
{\,},
\end{equation}
where along the way we have written 
$X^{+}/X^{-} = \sigK |X^{+}/X^{-}|$ [cf.~after 
(\ref{ratioXplusXminus})] and have used 
${\sf P}_{\xi} = - G$. Likewise, from (\ref{WKidentities}b) we find
\begin{equation}
G_{\pm} - \sigK e^{\mp 2\xi} G_{\mp} = \mp X^{\mp} {\sf P}_{X}
{\,}. 
\end{equation} 
With previous two equations and (\ref{cdcoords}a) we obtain 
the following new expressions
for the constraints:
\begin{eqnarray}
G_{\pm} & = & \Delta_{\pm}
  \left[V {\sf P}_{\xi} - \kappa \exp( -\alpha X) C' \mp
  2 \exp (-\alpha X) (C - W) {\sf P}_{X}\right] 
\label{newconstraints0} \eqnum{\ref{newconstraints0}a} \\
G & = & - {\sf P}_{\xi} {\,} ,
\eqnum{\ref{newconstraints0}b}
\addtocounter{equation}{1}
\end{eqnarray}
where the pre-factors are given 
by [cf.~Eq.~(\ref{ratioXplusXminus})]
\begin{equation}
\Delta_{\pm} := {\textstyle \frac{1}{2}} (X^{\pm})^{-1}
= - \sigma_{\pm}\sigL\frac{\exp ({\textstyle \frac{1}{2}}
\alpha X \mp \xi)}{\sqrt{4|C - w|}}
{\,} ,
\label{prefactor2}
\end{equation}
with $\sigma_{\pm}$ determined by 
$X^{\pm} = - \sigma_{\pm}\sigL |X^{\pm}|$ as before in $\S$ II. 
From (\ref{Killingnorm}) and
(\ref{cdcoords}a), it follows that $\sigma_{+}\sigma_{-} = \sigK$,
also as we have seen before.
The expression for $\bar{E}$ in (\ref{daggeraction}) is again
$\gamma E - v\gamma J$, but with the expressions for $E$ 
and $J$ given in terms of the new variables (\ref{cdcoords}).
One finds the new expressions to be
\begin{eqnarray}
E & = & - \Lambda^{-1}\kappa(\kappa X' - P_{\xi}) 
\label{newvariableEandJ} \eqnum{\ref{newvariableEandJ}a}\\
J & = & 2\Lambda^{-1}{\sf P}_{C}(C - W)
\eqnum{\ref{newvariableEandJ}b}
{\,} ,
\addtocounter{equation}{1}
\end{eqnarray}
where here one should consider $\Lambda$ as short-hand for 
\begin{equation}
  \Lambda = \kappa^{-1} \exp({\textstyle \frac{1}{2}}\alpha X)
  |C - W|^{-1/2}\left[\left|\kappa^{2}(X')^{2} - 4({\sf P}_{C})^{2}
  (C - W)^{2} + ({\sf P}_{\xi})^{2} - 2 \kappa X' {\sf P}_{\xi}
  \right|\right]^{1/2}
{\,} .
\label{newvariableLambda}
\end{equation}
The derivation of this expression for $\Lambda$ requires inversion of 
the canonical transformation (\ref{cdcoords}). We shall not need
the expressions (\ref{newvariableEandJ}) and (\ref{newvariableLambda}). 
We have collected these
equations only to make the following two points: (i) there is a 
${\sf P}_{C}$ ``buried'' in the new variable expression for 
$\bar{E}$ (coming from the new-variable expression for $\Lambda$);
and (ii) the variable $\xi$ appears nowhere in the action 
$L_{\dagger}$ except in the kinetic term. Our point (ii) implies 
that the equation of motion for $\xi$'s conjugate is 
$\dot{\sf P}_{\xi} = 0$ (of course, $-{\sf P}_{\xi}$ is the Lorentz
constraint!); therefore, as expected, that the pair  
$\{{\sf P}_{\xi},\xi\}$ is indeed ``pure-gauge.''


\subsection{Canonical reduction}

In this subsection we turn to the canonical reduction of the 
(vacuum) {\sc ecm}s and {\sc gdt}s we study in this paper. 
That is to say, we 
shall solve all of the constraints and substitute their solutions
back into the action, thereby obtaining a reduced action for the 
``true degrees of freedom.'' Moreover, the variational principle 
associated with our reduced action will feature fixation of the 
same relevant geometric quantities on the boundary as those which 
are fixed in the original action principle. 

Inspection of (\ref{newconstraints0}) shows that the vanishing of 
the set $\{{\sf P}_{X}, {\sf P}_{\xi}, C'\}$ is completely 
equivalent to the vanishing of the constraints. Therefore, let us 
re-express the action (\ref{daggeraction}) as
\begin{equation}
  L_{\dagger} 
= \int_{\cal M}{\rm d}^{2}x(
  {\sf P}_{C} \dot{C} 
+ {\sf P}_{X} \dot {X} 
+ {\sf P}_{\xi} \dot{\xi} 
- N^{C} C' - N^{X}{\sf P}_{X} 
- N^{\xi} {\sf P}_{\xi})
- \int_{\bar{\cal T}} {\rm d}t
  [\kappa (\eta - \hat{\eta}) 
  \dot{X}
+ \bar{N} \bar{E}]
{\, } ,
\label{newvariableaction}
\end{equation}
with 
\begin{eqnarray}
  N^{C} & = & 
- \kappa \exp( -\alpha X)
  (\Delta_{+} e^{+}{}_{t} 
+ \Delta_{-} e^{-}{}_{t})
\label{newlagrange} 
\eqnum{\ref{newlagrange}a} \\
  N^{X} & = & 
  2 \exp (-\alpha X) (C - W) 
  (\Delta_{-} e^{-}{}_{t}
- \Delta_{+} e^{+}{}_{t} )
\eqnum{\ref{newlagrange}b} \\
  N^{\xi} & = & 
  \omega_{t} 
- \kappa V( \Delta_{+} e^{+}{}_{t} 
+ \Delta_{-} e^{-}{}_{t}) 
{\, }. 
\eqnum{\ref{newlagrange}c} 
\addtocounter{equation}{1}
\end{eqnarray}
The next step is to reinterpret the expressions (\ref{newlagrange})
as freely variable Lagrange multipliers. However, as taking this 
step requires extreme caution, let us first examine the geometric 
interpretation of these multipliers in terms of the old variables. 
Using the definitions of $\bar{E}$ and $\bar{J}$, (\ref{cdcoords}a),
(\ref{prefactor2}), and the 
identities (\ref{geometricidentity}a,b), we find that
\begin{eqnarray}
  N^{C} & = & - {\textstyle \frac{1}{2}}(C - W)^{-1} 
  \bar{N} \bar{E}
\eqnum{\ref{newlapseandshift}a} \\
  N^{X}  & = & 
  \kappa^{-1} \bar{N} \bar{J}{\,} .
\label{newlapseandshift}
\eqnum{\ref{newlapseandshift}b}
\addtocounter{equation}{1}
\end{eqnarray}
Note that (\ref{newlapseandshift}b) implies the {\em on-shell} 
equality $N^{X} = \dot{X}$. We shall find these expressions useful
even after we have assumed that $N^{C}$ and $N^{X}$ are freely 
variable, an assumption which requires careful handling of the
boundary terms in the action.

One of the new constraints is $C'$; therefore, upon elimination of
this constraint one expects the emergence of a new canonical pair 
$\{{\bf P}_{\bf C}, {\bf C}\}$ defined by 
${\bf C}(t) := C(t)$ (the remaining ``zero-mode'' of $C$) 
and ${\bf P}_{\bf C} := \int_{\Sigma}{\rm d}r {\sf P}_{C}$. 
Moreover, one would like to
express the reduced action directly in terms of 
$\{{\bf P}_{\bf C},{\bf C}\}$.
However, note that our point (i) made after (\ref{newvariableLambda})
would seem to obstruct the fulfillment of this wish, as the momentum 
${\sf P}_{C}$ lies in boundary term explicitly. To bypass this 
obstruction and obtain the ``correct'' new action principle with
freely variable $N^{C}$, $N^{X}$, and $N^{\xi}$, we shall follow the 
prescription explained in Refs.~\cite{Louko_Whiting,Lau}
(which examined only cases of vacuum {\sc ssgr} and {\footnotesize 2d}{\sc dg}). Let us first
go through the prescription to obtain the new action, and only 
afterwards comment on why this new action principle is the ``correct''
one.

The first ingredient we need is the identity
\begin{equation}
 \bar{N}^{2} 
= \exp(\alpha X)\left[2 \kappa^{-2}(C - W) (N^{C})^{2} 
- {\textstyle \frac{1}{2}}(C - W)^{-1} (N^{X})^{2}\right]
{\, } ,
\label{newNbarsquared}
\end{equation}
which may be verified by explicit calculation with 
(\ref{newlagrange}a,b). Henceforth, the boldface symbol $\bbN$ 
shall represent the square root of the right-hand side of 
(\ref{newNbarsquared}). Next, using (\ref{QLE}) and (\ref{J}), we 
find that 
\begin{equation}
  E^{2} - J^{2} 
= 2 \kappa^{2} X^{+} X^{-}
{\,} .
\end{equation}
As is evident, the ``boost invariant'' $E^{2} - J^{2}$ is 
positive in static regions and negative in dynamical ones. By the 
relations (\ref{boostedQLE}), we know that $E^{2} - J^{2}
= \bar{E}^{2} - \bar{J}^{2}$; and, therefore, using (\ref{cdcoords}a),
we have
\begin{equation} 
  \bar{E}^{2} 
=  2 \kappa^{2} \exp(-\alpha X) (C - W) 
+ \bar{J}^{2}
{\,} .
\end{equation}
At this point, we use the {\sc eom} (\ref{eom}b) and the definition 
(\ref{Jbar}) to define
\begin{equation} 
\bbJ := 
\kappa \dot{X}/\bbN
{\,} ,
\label{newmomentum} 
\end{equation}
which agrees with  $\bar{J}$ on-shell. It now follows that, also 
on-shell, the expression we have considered for $\bar{E}$ is 
equivalent to 
\begin{equation}
  \bbE := - \kappa \sqrt{2\exp(-\alpha X) (C - W) 
+ (\dot{X}/\bbN)^{2}}
{\,} .
\label{newenergy}
\end{equation}
Notice that we have taken the negative square root. In fact, this is 
the appropriate sign choice for the (outer boundary-point) energy 
corresponding to our type I region [analogous to the (exterior) 
region I of the Penrose diagram for the {\sc sbh}]. The positive square root 
is the appropriate choice for the (outer boundary-point) energy 
expression corresponding to a type III region in the same 
example.\footnote{In effect, the 
roles of ``outer'' and ``inner'' are reversed in a type III region. We 
have assumed that our coordinate $r$ increases monotonically from right 
to left in a Kruskal-type diagram.} In type II and IV regions the sign 
of the energy expression need not be everywhere the same (and, 
therefore, can be zero). The prescription we shall follow seems to 
be well-suited only when the energy expression takes a definite 
sign, i.~e.~within the 
$\bar{\cal T}$ boundary lies entirely within static regions. 
There are two possibilities: (i) both the timelike boundary 
elements $\bar{\cal T}_{i}$ and $\bar{\cal T}_{o}$ lie within 
the same static region (or, in other words, the $\Sigma$ slices 
nowhere cut across the horizon), and (ii) the boundary 
elements $\bar{T}_{i}$ and $\bar{\cal T}_{o}$ lie in separate 
static regions (so that the $\Sigma$ slices do cut across the 
horizon and penetrate dynamical regions). Of course, possibility
(ii) is perhaps the more interesting one, as it allows one to 
study foliations which sample larger portions of the maximal
extension of $g_{\mu\nu}$. However, due caution is required 
when such a possibility is considered, as the canonical variables
(\ref{cdcoords}) would seem to be ill-defined at the horizon.
For the case of {\sc ssgr} Kucha\v{r} has addressed in detail the 
situation in which the $\Sigma$ slices do cut completely across 
the horizon\cite{Kuchar}. 
Nevertheless, it is our understanding that with the
canonical variables (\ref{cdcoords}) this 
situation is
not completely understood; hence, we shall steer clear of this
issue. Besides our goal is to obtain an expression for the 
rest-frame energy, and for this goal the above possibility (i) 
suffices. Therefore, for the rest of the paper, we shall assume
that the patch ${\cal M}$ lies entirely within a single static
region (which for the sake of definiteness we take to be
a type I region).

The parameter $\eta - \hat{\eta}$ describes the boost between 
the boundary $\bar{\Sigma}$ and Killing $\hat{\Sigma}$ frames.
Therefore, we can combine the boost relations (\ref{boostedQLE}) 
and (\ref{boostedQLE2}) in order to show that
\begin{eqnarray}
  \eta - \hat{\eta} & = & 
  \boldeta := 
- {\textstyle \frac{1}{2}} \log
  \left|
  \frac{\bbE + \bbJ}{\bbE - \bbJ}
  \right| {\,} . \\
  {\rm on-shell\,\,equality} 
  & \uparrow & 
\end{eqnarray}
We now use the new expressions 
(\ref{newenergy}) and (\ref{newmomentum}) to define a slightly 
different version of the action (\ref{newvariableaction}),
\begin{equation}
L_{\ddagger} =   
  \int_{\cal M}{\rm d}^{2} x 
  ({\sf P}_{C}\dot{C} 
+ {\sf P}_{X}\dot{X} 
+ {\sf P}_{\xi} \dot{\xi} - N^{C} C' 
- N^{X} {\sf P}_{X} 
- N^{\xi} {\sf P}_{\xi})
- \int_{\bar{\cal T}} {\rm d}t 
  \left(\kappa \boldeta \dot{X} + 
  \bbN \bbE \right)
{\,} .
\label{newvariableaction2}
\end{equation}
With $L_{\ddagger}$ we assume that the new Lagrange multipliers are 
freely variable. Let us verify that the new action $L_{\ddagger}$ 
possesses essentially the same variational principle as the original one 
$L$ from (\ref{canonicalaction1}). First, it is easy to find the 
constraints and canonical equations of motion associated with
$L_{\ddagger}$. These are the following:
\begin{eqnarray}
C' & = & 0
\label{canonicaleoms} 
\eqnum{\ref{canonicaleoms}a} \\
{\sf P}_{X} & = & 0
\eqnum{\ref{canonicaleoms}b} \\
{\sf P}_{\xi} & = & 0
\eqnum{\ref{canonicaleoms}c} \\
\dot{{\sf P}}_{C} & = & (N^{C})'
\eqnum{\ref{canonicaleoms}d} \\
\dot{C} & = & 0 
\eqnum{\ref{canonicaleoms}e} \\
\dot{{\sf P}}_{X} & = & 0
\eqnum{\ref{canonicaleoms}f} \\
\dot{X} & = & N^{X}
\eqnum{\ref{canonicaleoms}g} \\
\dot{{\sf P}}_{\xi} & = & 0
\eqnum{\ref{canonicaleoms}h} \\
\dot{\xi} & = & N^{\xi} {\,} .
\eqnum{\ref{canonicaleoms}i} 
\addtocounter{equation}{1}
\end{eqnarray}
Recall that with (\ref{newlapseandshift}b) we have already 
found (\ref{canonicaleoms}g) to be a valid equation on-shell. 
Notice that both $\{{\sf P}_{X},X\}$ and $\{{\sf P}_{\xi},\xi\}$ 
are ``pure-gauge'' pairs. 

The next step is to examine the boundary terms 
$(\delta L_{\ddagger})_{\partial {\cal M}}$ in the variation 
$\delta L_{\ddagger}$ of the new action. By a tedious but 
straightforward calculation we find
\begin{eqnarray}
  (\delta L_{\ddagger})_{\partial {\cal M}}  & = &  
  \int^{\Sigma''}_{\Sigma'} {\rm d}r 
  ({\sf P}_{C} \delta C 
+ {\sf P}_{X} \delta X  
+ {\sf P}_{\xi} \delta \xi) 
\nonumber \\
& & 
+ \int_{\bar{\cal T}} {\rm d}t 
   (\bbPi_{\bar{N}}\delta \bbN 
+ \bbPi_{C}\delta C 
+ \bbPi_{X} \delta X) 
- \left. \kappa {\boldeta} 
  \delta X \right|^{B''}_{B'} 
{\,} ,
\end{eqnarray}
with the following new $\bar{\cal T}$ momenta:
\begin{eqnarray}
  \bbPi_{\bar{N}} & = &
- \bbE
\label{reducedmomenta} 
\eqnum{\ref{reducedmomenta}a} \\
  \bbPi_{X} & = & 
  \bbN\left\{ 
  {\textstyle \frac{1}{2}}\bbE\left[
  \alpha + (C - W)^{-1} \lambda^{2} V_{0}\right] 
+ \bar{u}[{\boldeta}]\right\}
\eqnum{\ref{reducedmomenta}b} \\
  \bbPi_{C} & = & 
- {\textstyle \frac{1}{2}} 
  (C - W)^{-1}
  \bbN\bbE 
- N^{C}
{\,} .
\eqnum{\ref{reducedmomenta}c} 
\addtocounter{equation}{1}
\end{eqnarray}
Of course, the momentum $\bbPi_{\bar{N}}$ 
is not quite the momentum $\bar{\Pi}_{\bar{N}}$ 
considered in (\ref{momenta1}c), but it does agree 
with $\bar{\Pi}_{\bar{N}}$ on-shell. For positive $N^{C}$ the 
momenta $\bbPi_{C}$ 
vanishes on-shell. This can be shown by inserting into
$\bbPi_{C}$ the full 
expression (\ref{newNbarsquared}) for 
$\bbN^{2}$, using the 
canonical equation of motion (\ref{canonicaleoms}g), and 
realizing that $(C - W)$ is positive in a static region as we 
have here. 
Moreover, it is natural to assume that $N^{C}$ is positive,
because as a canonical expression in terms of the old variables $C'$
generates evolution forward in time along orbits of the isometry in
our type I region of 
interest\cite{new_Strobl,Kuchar,Schaller_Strobl,Kloesch_Strobl}. 
Alternatively, one can ``cheat'' and use
the interpretation (\ref{newlapseandshift}a) to establish the result. 
Therefore, $C$
need not be fixed on the $\bar{\cal T}$ boundary in the variational
principle associated with $L_{\ddagger}$, as the equations of
motion ensure that $\bbPi_{C}$ vanishes for arbitrary
variations $\delta C$ about a classical solution. Hence, the variational
principle associated with $L_{\ddagger}$ is qualitatively similar to
the one associated with the original action $L$. Both variational
principles feature fixation only of $X$ and the metric data 
$\bbN$ on the $\bar{\cal T}$ 
boundary.\footnote{As is well-known, the boundary lapse is 
intimately related to the inverse Hawking temperature\cite{Hawkingtemp} 
(blue-shifted from infinity). Therefore, with these 
quantities fixed at the boundary, our action is closely related 
to the so-called thermodynamical action appropriate for examining 
black holes in the canonical ensemble.\cite{BMWY} This 
consideration further motivates retention of these quantities 
as fixed data on the $\bar{\cal T}$ boundary.} With regard to 
the boundary data fixed on $\Sigma'$ and $\Sigma''$, these two 
variational principles do, of course, differ. This difference stems 
from the different choice of $\Sigma$ canonical variables used for 
each.\footnote{With regard to the aforementioned thermodynamical 
scenario, note that such a difference is irrelevant. In such a
scenario, the initial $\Sigma'$ and final $\Sigma''$ spacelike 
slices are typically identified (leaving {\em no} temporal boundary), 
as inverse temperature corresponds to {\em periodic} imaginary time.}
Having established that $L_{\ddagger}$ possesses the correct
variational principle, it is now a trivial matter to eliminate the
constraints in order to find, in full detail,
\begin{eqnarray}
L_{\ddagger}|^{\rm red} & = &   
\int^{t''}_{t'}{\rm d}t {\,} {\bf P}_{\bf C}\dot{\bf C}
+ \int_{\bar{\cal T}} {\rm d}t 
  \left[ \sqrt{2\exp(-\alpha X)
  \bbN^{2} ({\bf C} - W) 
+ (\dot{X})^{2}}\right. \nonumber \\
& & 
\left. - {\textstyle \frac{1}{2}} 
  \kappa \dot{X}\log \left|
  \frac{\sqrt{2\exp(-\alpha X)
  \bbN^{2} ({\bf C} - W) 
+ (\dot{X})^{2}}
+ \dot{X} }{\sqrt{2\exp(-\alpha X)
  \bbN^{2}({\bf C}
- W) 
+ (\dot{X})^{2}}
- \dot{X}}\right|  \right]
\label{reducedaction}
\end{eqnarray}
as the promised reduced action. In the reduced action the boundary
lapse $\bbN$ is a positive independent parameter. 
This is the reduced
action corresponding to a {\em general} {\sc ecm} for our 
bounded patch ${\cal M}$.

\subsection{Rest-frame energy}
Let us now derive an expression for the rest-frame energy 
which is applicable to those {\sc gdt}s which have an 
asymptotically flat black-hole solution and a flat-space 
``linear dilaton'' vacuum. Namely, those models within the
physical class of $\S$ IV. Again, we continue to 
assume the vacuum case, although all we really need to require 
is that matter, if present, has support on a compact set which 
lies within the outer boundary $\bar{\cal T}_{o}$.  The scenario 
we study now has been outlined in $\S$ IV. Again, without loss 
of generality, we set $\alpha = 0$. First, performance of
the conformal transformation of $\S$ I.C in the action 
$L_{\ddagger}$ amounts to a simply re-definition,
\begin{equation}
  \tilde{L}_{\ddagger} =   
  \int_{\cal M}{\rm d}^{2} x 
  \left({\sf P}_{C}\dot{C} 
+ {\sf P}_{X}\dot{X} 
+ {\sf P}_{\xi} \dot{\xi} 
- N^{C} C' - N^{X} {\sf P}_{X} 
- N^{\xi} {\sf P}_{\xi}\right)
- \int_{\bar{\cal T}} {\rm d}t 
  \left(\kappa \boldeta \dot{X} + \tilde{\bbN}
  \tilde{\bbE} \right)
{\,} ,
\end{equation}
with
\begin{eqnarray}
  \tilde{\bbN}{}^{2} & = & 
  \exp(- 2\varphi)
  \left[2 \kappa^{-2}(C - W)
  (N^{C})^{2} 
- {\textstyle \frac{1}{2}}(C - W)^{-1} 
 (N^{X})^{2}\right]
\label{newtildeNandE} 
\eqnum{\ref{newtildeNandE}a} \\
  \tilde{\bbE} & = & - \kappa 
  \sqrt{2\exp(2\varphi) (C - W) 
+ \left(\dot{X}/
  \tilde{\bbN}\right)^{2}}
\eqnum{\ref{newtildeNandE}b} \\
\boldeta & = & 
- {\textstyle \frac{1}{2}} \log
  \left|
  \frac{\tilde{\bbN} \tilde{\bbE} + \kappa \dot{X}}{
        \tilde{\bbN}\tilde{\bbE}  - \kappa \dot{X}}
  \right| {\,} .
\eqnum{\ref{newtildeNandE}c}
\addtocounter{equation}{1}
\end{eqnarray}
The {\sc eom}s associated with $\tilde{L}_{\ddagger}$ are the same 
ones (\ref{canonicaleoms}) as before; and, moreover, the 
variational principle associated with $\tilde{L}_{\ddagger}$ 
only features fixation of $X$ and 
$\tilde{\bbN}$ on the 
$\bar{\cal T}$ boundary (and is identical in important respects 
to the one associated with $L_{\ddagger}$).

Next, we shall assume that the variational set of histories associated 
with the action $\tilde{L}_{\ddagger}$ is determined by a fixed 
value of $X$ on the $\bar{\cal T}$ boundary, and so $\dot{X} = 0$. 
This in turn implies that on-shell
the quasilocal momentum $\tilde{\bar{J}{\,\,}}\!\! 
= \tilde{\bbJ\,\,}\!\! 
= 0$, which is why we are using the term ``rest frame.''  
Furthermore, this assumption 
implies that the $\bar{\cal T}$ boundary is generated by the flow 
of the Killing field $k^{\mu}$, or, in other words, that the 
frames $(\tilde{\bar{u}}{}^{\mu}, 
\tilde{\bar{n}}{}^{\mu})$ and 
$(\tilde{\hat{u}}{}^{\mu}, 
\tilde{\hat{n}}{}^{\mu})$ 
have been identified at the $\bar{\cal T}$ boundary. 
This rest-frame condition may be achieved 
precisely because we are working with an action principle which 
features fixation of $X$ as boundary data. The action now takes
 the form
\begin{eqnarray}
\tilde{L}_{\ddagger} & = &  
  \int_{\cal M}{\rm d}^{2} x \left({\sf P}_{C}\dot{C} 
+ {\sf P}_{X}\dot{X} + {\sf P}_{\xi} \dot{\xi} - N^{C} C' 
- N^{X} {\sf P}_{X} - N^{\xi} {\sf P}_{\xi}\right)
\nonumber \\
& & \int_{\bar{\cal T}} {\rm d}t 
  \left[2\kappa \lambda^{-1}
\tilde{\bbN}
  |W|\sqrt{1 + C|W|^{-1}}
\right]
{\,} ,
\end{eqnarray}
where we have chosen $\S$ IV's conformal transformation and assumed
those conditions relevant for the class of physical models, $W = -|W|$ 
and $\lim_{X\to\infty} |W| = \infty$.

We subtract from the action a 
{\em reference term}\cite{Brown_York}[cf.~Eq.~(\ref{PSMaction})]
\begin{equation}
\tilde{L}_{\ddagger}|^{\scriptscriptstyle 0} := 
\int_{\bar{\cal T}} {\rm d}t 
  \left[2\kappa \lambda^{-1}
\tilde{\bbN}
  |W|\right].
\label{referenceddagger}
\end{equation}
Notice that the reference term is a functional only of boundary 
data which is fixed in the variational principle. 
Physically, we may interpret  
$\tilde{L}_{\ddagger}|^{\scriptscriptstyle 0}$ as the action 
$\tilde{L}_{\ddagger}$
evaluated on the {\em linear dilaton vacuum} of $\S$ IV. More 
precisely, we consider the isometric 
embedding of the $\bar{\cal T}$ boundary 
in flat Minkowski spacetime as the (disjoint) union of 
inner an outer lines of 
constant $r = \lambda^{-1} \tilde{X}_{0}$ 
[cf.~Eq.~(\ref{tildeXnaught})]. Here $\tilde{X}_{0}$ is 
determined by the fixed value of $X$ on either 
$\bar{\cal T}_{i}$ or $\bar{\cal T}_{o}$, as in general 
$\tilde{X} = \tilde{X}(X)$. One should re-scale the inertial time 
coordinate in $({\rm d}\tilde{s}_{0})^{2}$ accordingly, in order
that, with respect to the rescaled time, 
$- \tilde{\bbN}{}^{2}$ 
is the induced metric on the these timelike lines in Minkowski 
spacetime. 
Since one is always free to add functionals of the fixed boundary 
data to the action without affecting the variational principle, 
we are assured of the fact that the variational principle 
associated with the new referenced action, 
\begin{eqnarray}
\tilde{L}_{\ddagger}|^{\scriptscriptstyle {\rm ref}}
:= \tilde{L}_{\ddagger} 
- \tilde{L}_{\ddagger}|^{\scriptscriptstyle 0} 
& = &  \int_{\cal M}{\rm d}^{2} x ({\sf P}_{C}\dot{C} 
+ {\sf P}_{X}\dot{X} + {\sf P}_{\xi} \dot{\xi} - N^{C} C' 
- N^{X} {\sf P}_{X} - N^{\xi} {\sf P}_{\xi})
\nonumber \\
& & - \int_{\bar{\cal T}} {\rm d}t 
  \left[2\kappa\lambda^{-1}
\tilde{\bbN}
  |W|\left(1 - \sqrt{1 + C|W|^{-1}}\right)
\right]
{\,},
\label{referencedaction}
\end{eqnarray}
is the same as before. As with (\ref{newvariableaction2}) above, 
elimination of the constraints in (\ref{referencedaction}) in order 
to find the corresponding reduced action is a trivial matter.

The (outer boundary-point) {\em referenced} quasilocal energy we 
``read off'' from the reduced form of the canonical action 
(\ref{referencedaction}) is the following one:
\begin{equation}
\tilde{\bbE}|^{\scriptscriptstyle {\rm ref}} = 
  2\kappa\lambda^{-1}
  |W|\left[1 - \sqrt{1 + {\bf C}|W|^{-1}}\right]
{\,} .
\end{equation}
This expression for the rest-frame {\sc qle} agrees with the 
result obtained in Ref.~\cite{Barvinsky_Kunstatter} by Barvinsky 
and Kunstatter directly from the Hamiltonian for a {\sc gdt} 
(taking $C'$ in place of the usual Hamiltonian constraint) 
via a Regge-Teitleboim-type argument\cite{Regge_Teitleboim}. 
That is, agreement is obtained, provided that one chooses a unit 
boundary lapse for the expression in that reference (as is 
appropriate, since the {\sc qle} is the Hamiltonian value 
corresponding to a pure time translation\cite{Brown_York}).
The $X \to \infty$ limit of 
$\tilde{\bbE}|^{\scriptscriptstyle {\rm ref}}$ 
defines the analog of the {\sc adm} mass 
(asymptotic rest-frame energy) for the models we 
consider\cite{Brown_York,Lau}. We find
\begin{equation}
M_{\scriptscriptstyle ADM} \equiv 
\lim_{X \to \infty} 
\tilde{\bbE}|^{\scriptscriptstyle {\rm ref}} = 
- \kappa\lambda^{-1} {\bf C}
{\,} .
\end{equation}
With the {\sc ssgr} choice (\ref{SSGRchoices}a) for $\kappa$,
$M_{\scriptscriptstyle ADM}$ is the Schwarzschild mass parameter
$M_{\scriptscriptstyle S}$ 
(as expressed in terms of ${\bf C} = C$ as given in $\S$ III and IV).
More generally, taking $\kappa = 1$ (as is perhaps more appropriate
for a 2d model), we obtain $M_{\scriptscriptstyle ADM} 
= - \lambda^{-1} {\bf C}$ (mass dimension 1 in 2d). Finally,
we mention that the quasilocal approach employed here may also be used 
to define total gravitational energy-momentum at null 
infinity.\cite{Mueller-Kirsten_et_al,Brown_York_Lau}


\section{Conclusions}


Our main results in this paper concern the important 
role played by boundary conditions in the passage -via 
conformal transformation- between Einstein-Cartan 
models ({\sc ecm}s, described by a first-order zweibein 
action) and 
generalized dilaton theories ({\sc gdt}s, described by a metric 
action). Once again, we have been motivated to study this 
equivalence because, although it is {\sc gdt} which more 
nearly mimics physical features of 4d 
spherically symmetric general relativity ({\sc ssgr}) and 
thus provides more interesting 2d models of gravity, it is 
far more tractable mathematically to work within the first-order 
framework associated with {\sc ecm}s. As shown in $\S$ I, if 
the intent is to perform such a conformal re-definition of 
field variables in the first-order action for an {\sc ecm}, 
then a boundary term must be included in the definition of 
the action, or else the action's associated variational 
principle is not preserved under the the conformal 
transformation. Moreover, while at first sight it appears 
that the addition of such a boundary term destroys the 
zweibein gauge-invariance of the theory at the boundary
$\partial {\cal M}$ of the spacetime patch ${\cal M}$, we 
have explicitly demonstrated in $\S$ V that this need not 
be true. It is quite notable that by imposing the dual 
requirements of (i) conformally preserved boundary conditions 
and (ii) complete zweibein-gauge invariance, we have arrived 
at set of canonical variables, namely, the $\Sigma$-adapted 
canonical variables ($\Sigma${\sc cv}s) (\ref{Sigmavariables}), 
which appear to be new in the literature. Although 
strictly zweibein variables, the $\Sigma${\sc cv}s are 
intimately related to the standard canonical {\sc adm} 
metric variables, and therefore, they particularly elucidate 
the relationship between zweibein {\sc ecm}s and 
metric {\sc gdt}s. Moreover, as we discuss below, the 
$\Sigma${\sc cv}s also play an important part in the overall 
picture of the inter-relationships between several
author's approaches to 1+1 canonical gravity. 

We have also throughly explored the notion of gravitational 
energy and momentum in 1+1 gravity. Besides being of 
interest in its own right, the canonical quasilocal 
energy-momentum (here derived via a 
Hamilton-Jacobi-type argument due to Brown 
and York\cite{Brown_York}) 
serves as a stepping stone for introducing the rapidity 
in $\S$ V.B, thus leading to the rapidity canonical variables 
({\sc rcv}s) in $\S$ VI.B. 
With the $\Sigma${\sc cv}s and {\sc rcv}s in mind, 
let us return to the various canonical transformations 
examined in the last section. A ``flow chart'' in Fig.~2
depicts the inter-relationships between the various 
canonical variables we have introduced. At the top of the 
chart are the $\Sigma${\sc cv}s.
The chain of transformations labeled $1$, $2$, and $3$ 
correspond to passage from a zweibein {\sc ecm} to 
a metric {\sc gdt}. As we have noted in $\S$ VI.A, 
one could interchange the order of steps 2 and 3, 
i.~e.~the canonical transformation corresponding to 
the conformal rescaling can be performed prior to 
elimination of the gauge pair $\{P_{\rho},\rho\}$. 
Notice that in the 1-2-3 chain, no generating functionals 
are required. This further underscores the significance of the 
$\Sigma${\sc cv}s at the top of the chart and why we have called them
``$\Sigma$-adapted.'' Steps {\sc i} and {\sc a} depict the 
canonical transformation to the {\sc rcv}s (\ref{cdcoords}). 
Essentially the same generating functional, $\Xi$ given 
in Eq.~(\ref{generatingfunctionalfirst}), is required for 
both steps. From the {\sc rcv}s, one may easily pass to the 
``true degrees of freedom'' via the chain labeled by 
{\sc b-c}. {\em Not} depicted in our flow chart are the 
Possion-sigma-model canonical variables [cf.~the 
discussion immediately preceding Eq.~(\ref{identityone})] 
of Kl\"{o}sch and Strobl \cite{Kloesch_Strobl}, although as 
our main discussion indicates they lie half-way between 
the $\Sigma${\sc cv}s and the {\sc rcv}s (in the middle of 
the {\sc a} transformation). The canonical 
transformations {\sc ii} and {\sc iii}, made after 
elimination of the Lorentz constraint, are quite closely 
related to the canonical transformation made in {\sc i} 
(or in {\sc a}). In fact, the generating functionals 
$\Xi'$ and $\tilde{\Xi}'$ for these steps are defined 
from $\Xi$ by elimination of the Lorentz 
constraint,\footnote{Note that neither $\Xi'$ nor 
$\tilde{\Xi}'$ are the same as the alternative generating 
functional $\Upsilon$ considered in $\S$ VI. However, it is 
true that $\Xi' = \tilde{\Xi}' = \Upsilon|_{G=0}$. Of course, 
$\Upsilon \neq \Xi |_{G = 0}$.}
i.~e.~$\Xi' = \tilde{\Xi}' = \Xi|_{G=0}$. Clearly,
$\Xi'$ and $\tilde{\Xi}'$, although equal to each other, 
are expressed in terms of different canonical variables. 
In fact, one finds
\begin{eqnarray}
\Xi'|_{\Sigma} & := & - \int_{\Sigma} {\rm d}r
\left[\Lambda P_{\Lambda} + 
{\textstyle \frac{1}{2}}X'
                    \log\left|\frac{X' + \Lambda
                               P_{\Lambda}}{X' 
                              - \Lambda P_{\Lambda}}
                                \right|\right]
\nonumber \\
& = & 
   - \int_{\Sigma}{\rm d}r 
          \left[\tilde{\Lambda}{\cal P}_{\tilde{\Lambda}} + 
                {\textstyle \frac{1}{2}}X'
                    \log\left|\frac{X' + \tilde{\Lambda}
                              {\cal P}_{\tilde{\Lambda}}}{X' 
                              - \tilde{\Lambda}{\cal P}_{
                                \tilde{\Lambda}}}
                                \right|\right]
=: \tilde{\Xi}'|_{\Sigma}
{\,} . \nonumber
\end{eqnarray}
For the case of vacuum {\sc ssgr}, the generating functional 
$\tilde{\Xi}'$ is precisely the one considered 
by Kucha\v{r} in his thorough paper on the canonical 
geometrodynamics of {\sc sbh}s\cite{Kuchar}. This generating
functional also plays an important role in Louko and 
Whiting's examination (employing Kucha\v{r}'s approach)
of the Hamiltonian thermodynamics of 
{\sc sbh}s\cite{Louko_Whiting}. For vacuum 
{\footnotesize 2d}{\sc dg} it corresponds 
to the generating functional considered in Ref.~\cite{Lau}.
At the level of full {\sc gdt}, it is precisely such generating
functionals which correspond to the transformations studied by
Kunstatter and Barvinsky\cite{Barvinsky_Kunstatter}. 

Finally, let us discuss what we have {\em not} done in this 
paper. First, although we have striven to point out precisely where 
and why the inclusion of matter does not affect our results, we
have not taken matter explicitly into account. It would be 
interesting to examine matter-induced gravitational collapse in
{\sc gdt} along the same lines as carried out in {\sc ssgr} by
Guven and \'{O} Murchadha\cite{Guven_O_Murchadha}, especially as 
mathematically their formalism is quite similar to the one studied
here [cf.~the footnote just before 
Eq.~(\ref{constraints0})]. The known existence of critical behavior 
for gravitational collapse in {\sc ssgr} would seem to indicate that 
there are, in fact, a wide class of models where such criticality 
could be examined (here {\em non}-minimally coupled matter
would be a prerequisite). For treatments of the coupled 
matter-gravity {\sc cghs} model\cite{CGHS},
Refs.~\cite{SHayward,Kuchar_et_al} should be consulted.  
Second, we have bypassed the stubborn problem of either (i) 
finding good canonical variables\footnote{Of course, depending
on the temporal foliation, the plain {\sc adm} variables certainly
may be well-behaved across a horizon; but we seek variables 
which simplify (preferably drastically) the constraints as well. 
The {\sc adm} variables do not yield such simplification.} 
which do not break 
down on horizons or, at the very least, (ii) developing 
a precise understanding of how to use known 
``target-space'' canonical variables (such as our {\sc rcv}s) 
when a spacelike slice cuts completely across a horizon
in a general way. As mentioned, for {\sc ssgr} Kucha\v{r} has 
studied in detail option (ii)\cite{Kuchar}, and his treatment is 
the most serious attempt in this direction known to us. However, 
in our opinion this issue is quite deserving of further research.


\section*{Acknowledgments}


For helpful discussions and insight we thank H. Balasin, H. Liebl, 
N. \`{O} Murchadha, and, particularly, both T. Kl\"{o}sch and 
T. Strobl. This research has been supported by the 
``Fonds zur F\"{o}rderung 
der wissenschaftlichen Forschung'' in Austria (FWF project 
10.221-PHY). S.~R.~Lau also gratefully acknowledges support as a 
Lise Meitner Fellow of the FWF (Project M-00182-PHY).


\begin{figure}
\epsfxsize=5in
\centerline{\epsfbox{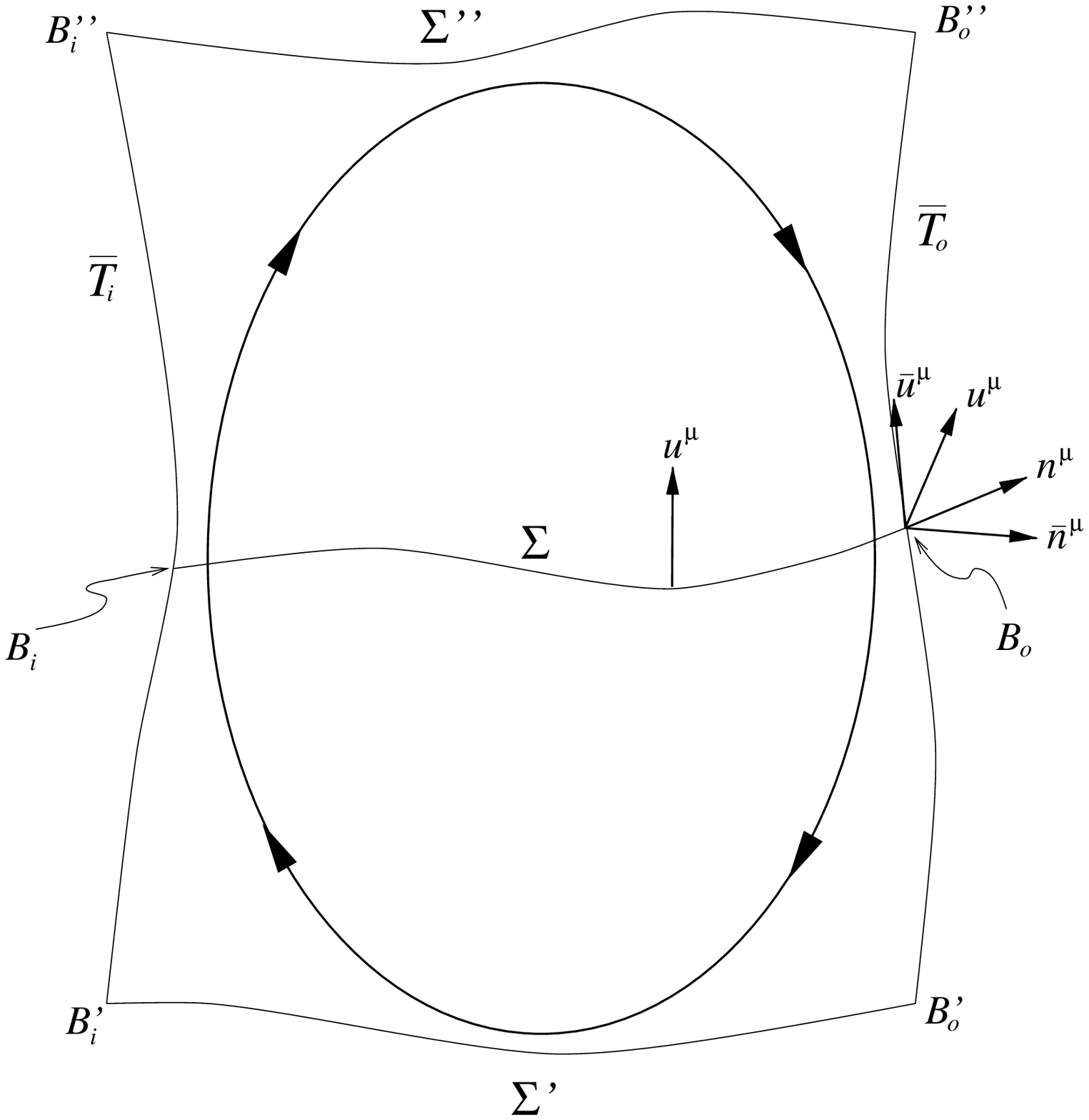}}
\caption{Spacetime Patch ${\cal M}$}
{\small 
The figure depicts the geometry of our spacetime 
patch ${\cal M}$. Here $\Sigma$ represents a 
generic $1$-dimensional spacelike slice
with a boundary consisting of two points, 
$B_{i}$ and $B_{o}$.
At the outer boundary point $B_{o}$ the time-gauge
zweibein $(u^{\mu}, n^{\mu})$ and the radial-gauge 
zweibein $(\bar{u}^{\mu}, \bar{n}^{\mu})$ are drawn. 
Note that $n^{\mu}$ points tangentially to $\Sigma$ 
and is the outward-pointing normal of $B_{o}$ as 
embedded in $\Sigma$. The vector $\bar{u}^{\mu}$ 
points tangentially to the $1$-dimensional timelike 
curve $\bar{\cal T}_{o}$ (one element of the full
boundary $\partial {\cal M}$) and is the 
future-pointing normal of $B_{o}$ as embedded in 
$\bar{\cal T}_{o}$. All the elements of 
$\partial {\cal M}$, namely, $\Sigma'$, $\Sigma''$,
$\bar{\cal T}_{i}$, and $\bar{\cal T}_{o}$, are 
depicted. Notice that 
$\eta = \sinh^{-1}( - u_{\mu} \bar{u}^{\mu})$
need not vanish at the outer boundary point $B_{o}$.
For that matter, the boost parameter $\eta$ need not 
vanish at the corner points: $B'_{i}$, $B''_{i}$,
$B'_{o}$, and $B''_{o}$.
The heavy ellipse inscribed within ${\cal M}$
illustrates our convention for the orientation of 
the boundary $\partial {\cal M}$ as embedded in 
${\cal M}$. The chosen ``clockwise'' orientation 
determines the integration conventions (\ref{orientation}).
}
\end{figure}
 \begin{figure}
\epsfxsize=6in
\centerline{\epsfbox{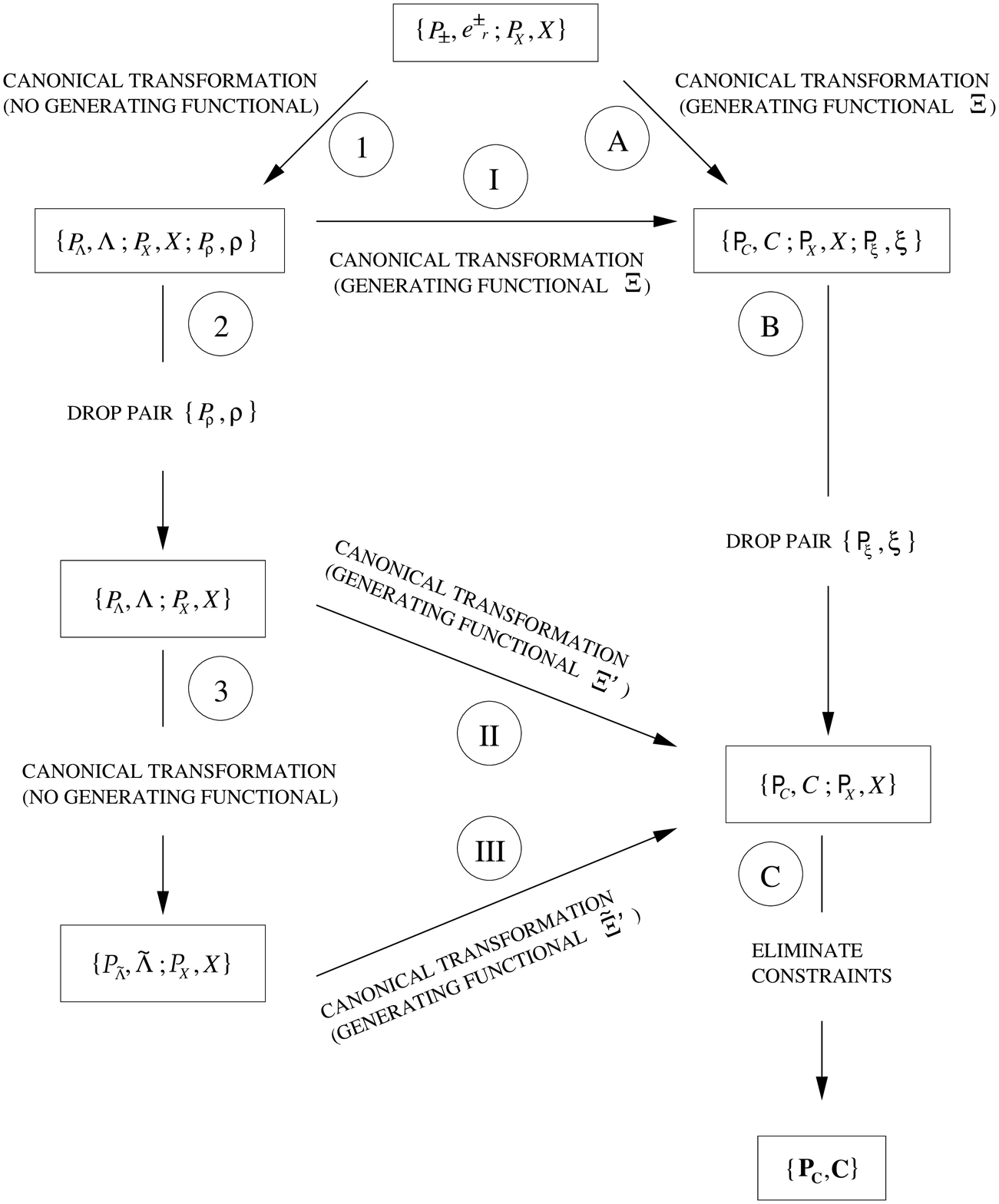}}
\caption{Canonical Variables and Transformations from $\S$ VI}
{\small The various canonical variables and canonical 
transformations depicted in this flow chart are described in full 
in $\S$ VII, the concluding section.}
\end{figure}
\end{document}